\documentclass[preprint]{elsarticle}

\usepackage{subcaption}
\usepackage{xspace}
\usepackage{multirow}
\usepackage[algo2e,ruled,vlined]{algorithm2e}
\usepackage{url}
\usepackage{amsmath}
\usepackage{booktabs}
\usepackage{rotating}
\usepackage{float}

\graphicspath{{./figures/}}

\newtheorem{exmp}{Example}[section]
\newcommand{\code}[1]{{\textnormal{\texttt{#1}}}}
\newcommand{\ppn}{\textnormal{\texttt{ppn}}\xspace}

 \newcommand{\locality}{\textnormal{\texttt{locality}}\xspace}

\begin{document}
\begin{frontmatter}
    \title{Node aware sparse matrix-vector multiplication\tnoteref{funding}}
    \tnotetext[funding]{This research is part of the Blue Waters sustained-petascale computing project,
which is supported by the National Science Foundation (awards OCI-0725070
and ACI-1238993) and the state of Illinois. Blue Waters is a joint effort of
the University of Illinois at Urbana-Champaign and its National Center for
Supercomputing Applications.  This material is based in part upon work
supported by the National Science Foundation Graduate Research Fellowship
Program under Grant Number DGE-1144245.  This material is based in part upon
work supported by the Department of Energy, National Nuclear Security
Administration, under Award Number DE-NA0002374. }

\author{Amanda Bienz, William D. Gropp, and Luke N. Olson}
    \address{Department of Computer Science\\
        University of Illinois at Urbana-Champaign\\
        Urbana, IL 61801}

\begin{abstract}
    The sparse matrix-vector multiply (SpMV) operation is a key computational
    kernel in many simulations and linear solvers.  The large communication
    requirements associated with a reference implementation of a parallel SpMV
    result in poor parallel scalability.  The cost of communication depends on
    the physical locations of the send and receive processes: messages injected
    into the network are more costly than messages sent between processes on
    the same node.  In this paper, a node aware parallel SpMV (NAPSpMV) is
    introduced to exploit knowledge of the system topology, specifically the
    node-processor layout, to reduce costs associated with communication.  The
    values of the input vector are redistributed to minimize both the number and
    the size of messages that are injected into the network during a SpMV,
    leading to a reduction in communication costs.  A variety of computational
    experiments that highlight the efficiency of this approach are presented.
\end{abstract}


\end{frontmatter}

\section{Introduction}\label{section:intro}
\begin{figure}[!ht]
    \centering
    \includegraphics[width=0.6\textwidth]{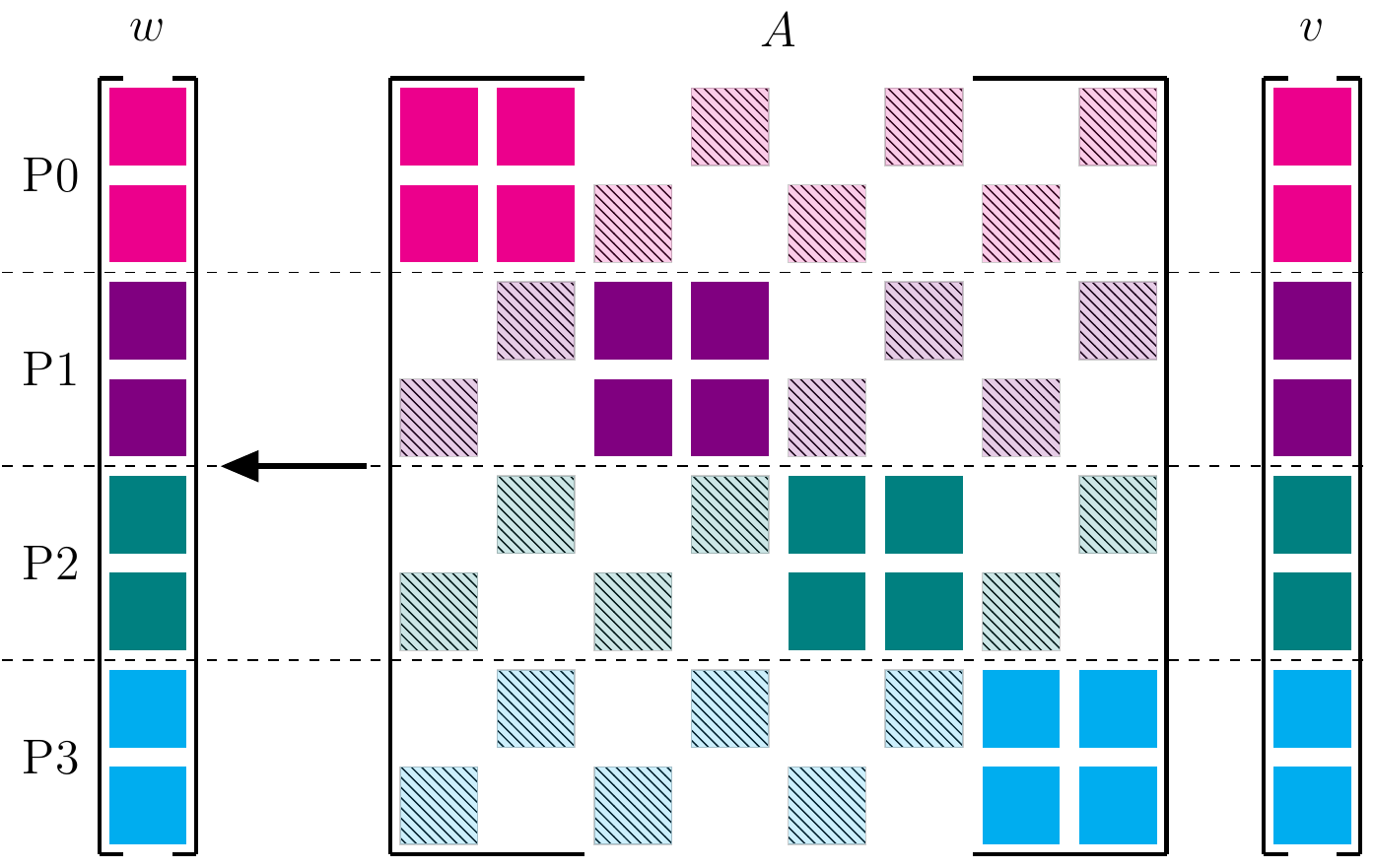}
    \caption{A matrix partitioned across four processes, where each
    process stores two rows of the matrix, and the equivalent rows of each
    vector.  The on-process block of each matrix partition is represented by solid
    squares, while the off-process block is represented by patterned
    entries.}\label{figure:partition}
\end{figure}

The sparse matrix-vector multiply (SpMV) is a widely used operation in many
simulations and the main kernel in iterative solvers.  The focus of this paper
is on the parallel SpMV, namely
\begin{equation}
w \leftarrow A \cdot v\\
\end{equation}
where $A$ is a sparse $N \times N$ matrix and $v$ is a dense $N$-dimensional
vector.  In parallel, the sparse system is often distributed across $n_{p}$
processes such that each process holds a contiguous block of rows from the
matrix $A$, and equivalent rows from the vectors $v$ and $w$, as shown in
Figure~\ref{figure:partition}.  A common approach is to also split the rows of
$A$ on a single process into two groups: an on-process block, containing the
columns of the matrix that correspond to vector values stored locally, and an
off-process block, containing matrix non-zeros that are associated with vector
values that are stored on non-local processes.  Therefore, non-zeros in the
off-process block of the matrix require vector values to be communicated during
each SpMV\@.

The SpMV operation lacks parallel scalability due to large costs associated with
communication, specifically in the strong scaling limit of a few rows per
process.  Increasing the number of processes that a matrix is distributed across
increases the number of columns in the off-process blocks, yielding a growth in
communication.

\begin{figure}[!ht]
    \centering
    \includegraphics[width=0.6\textwidth]{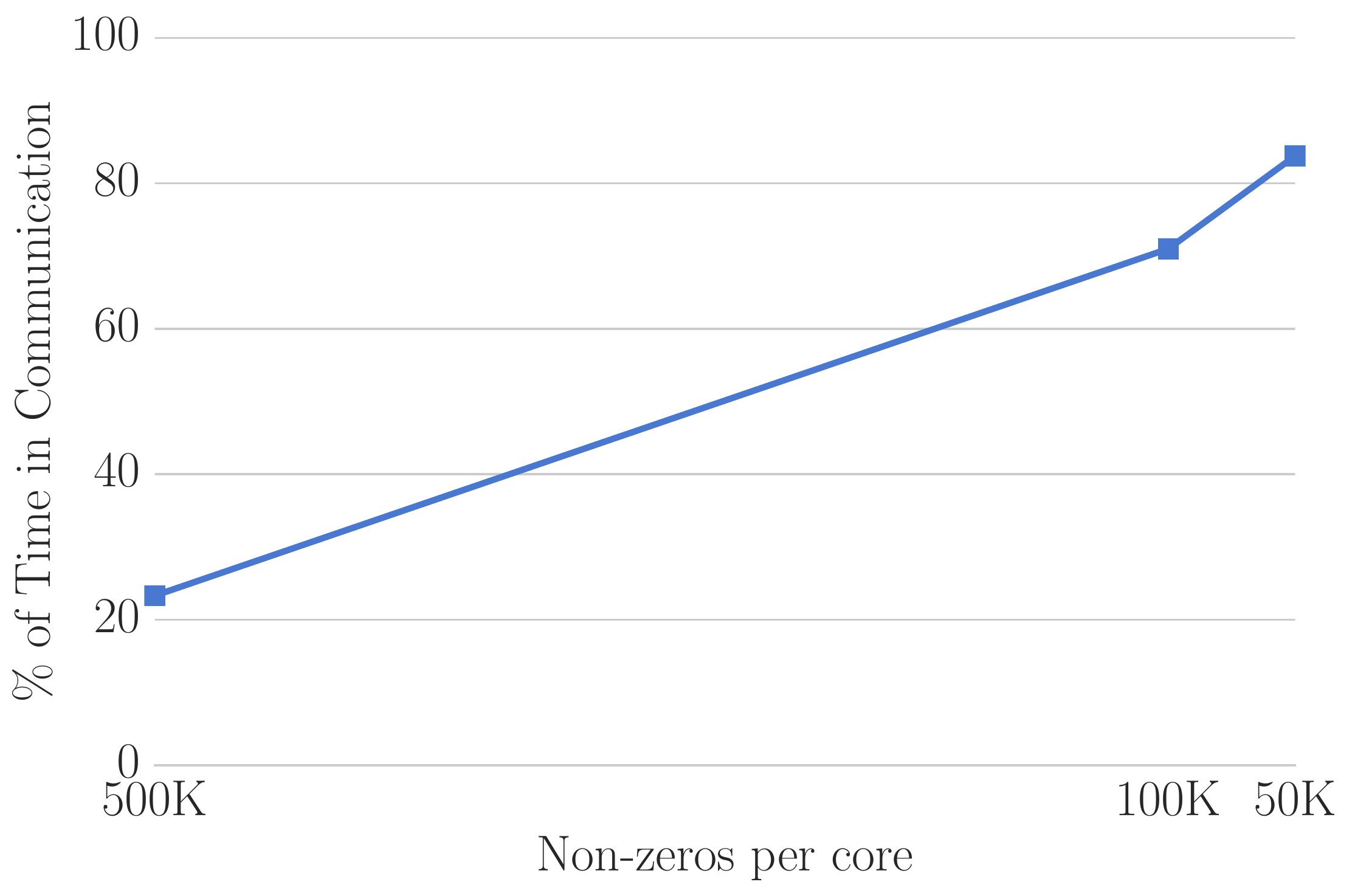}
    \caption{Percentage of total SpMV time spent during communication for
    matrix \code{nlpkkt240} with 760,648,352 non-zeros}\label{figure:ufl_spmv_times}
\end{figure}

Figure~\ref{figure:ufl_spmv_times} shows the percentage of time spent
communicating during a SpMV operation for two large matrices from the 
SuiteSparse matrix collection at scales varying from
$50\,000$ to $500\,000$ non-zeros per process~\cite{SuiteSparse}.  The results show that the
communication time dominates the computation as the number of processes
is increased, thus decreasing the scalability.

Machine topology plays an important role in the cost of
communication~\cite{Williams}.
Multicore distributed systems present new challenges in communication as the
bandwidth is limited while the number of cores participating in communication
increases~\cite{2016GrOlSa_pingpong}.  Injection limits and network contention
are significant roadblocks in the SpMV operation, motivating the need for SpMV
algorithms that take advantage of the machine topology.  The focus of the
approach developed in this paper is to use the node-processor hierarchy to more
efficiently map communication, leading to notable reductions in SpMV costs on
modern HPC systems for a range of sparse matrix patterns. Throughout 
this paper, the term \textit{node aware} refers to knowledge of the mapping of
processes to physical nodes, although other aspects of the topology~---~e.g.\
socket information~---~could be used in a similar fashion.  The mapping of
virtual ranks to physical processors can be easily determined on many super
computers.  The flag \code{MPICH\_RANK\_REORDER\_METHOD} can be set to a predetermined
ordering on Cray machines, while modern Blue Gene machines allow the user to specify
the ordering among the coordinates A, B, C, D, E, and T through the variable
\code{RUNJOB\_MAPPING} or a runscript option of \code{--mapping}.

There are a number of existing approaches for reducing communication costs
associated with sparse matrix-vector multiplication.  Communication volume in
particular is a limiting factor and the ordering and parallel partition of a
matrix both influence the total data volume.  In response, graph partitioning
techniques are used to identify more efficient layouts in the data~\cite{Pinar,
Hendrickson, BisselingDataDist, TwoDimPart}.
ParMETIS~\cite{1998_parmetis} and PT-Scotch~\cite{2008_ptscotch}, for example, provide parallel
partitioning of matrices that often lead to improved system loads and more
efficient sparse matrix operations.  Communication volume is accurately modeled
through the use of a hypergraph~\cite{HypergraphPart}.  As a result, hypergraph partitioning also leads
to a reduction in parallel communication requirements, albeit at a larger
one-time setup cost.  Topology-aware task mapping is used to accurately map
partitions to the allocated nodes of a supercomputer, reducing the overall cost
associated with communication~\cite{Jeannot, Agarwal, Sreepathi, Malik, Traff}.  
The approach introduced in this paper complements these
efforts by providing an additional level of optimization in handling
communication.

Topology-aware methods and aggregation of data are commmonly used to reduce
communication costs, particularly in collective operations~\cite{Solomonik,
Kielmann, Karonis, Sack}.  Aggregation of data is used in point to point
communication through Tram, a library for streamlining messages in which data
is aggregated and communicated only through neighboring processors~\cite{Tram}.
The method presented in this paper aggregates messages at the node level and
communicates all aggregated data at once, yielding little structural change
from standard MPI communication while reducing overall cost.

The performance of matrix operations can also be improved through the use of hybrid
architectures and accelerators, such a graphics processing units (GPUs).  The
throughput of GPUs allows for improved performance when memory access patterns
are optimized~\cite{BellSpMV, Dalton}.

Many preconditioners for iterative methods, such as algebraic multigrid, rely on
the SpMV as a dominant operation and therefore lack scalability due to large
communication costs.  A variety of methods exist for altering the preconditioning
algorithms to reduce the communication costs associated with each
SpMV~\cite{Sterck, Sparsify, BellAMG}.

This paper focuses on increasing the locality of communication during a SpMV to
reduce the amount of communication injected into the network.
Section~\ref{section:background} describes a reference algorithm for a parallel
SpMV, which resembles the approach commonly used in practice.  A performance
model is also introduced in Section~\ref{section:models}, which considers the
cost of intra- and inter-node communication and the impact on performance.  A
new SpMV algorithm is presented in Section~\ref{section:tapspmv}, which reduces
the number and size of inter-node messages by increasing the significantly
cheaper intra-node communication.  The code and numerics are presented in
Section~\ref{section:results} to verify the performance.

\section{Background}\label{section:background}
Modern supercomputers incorporate a large number of nodes through an
interconnect to form a multi-dimensional grid or torus network.  Standard
compute nodes are comprised of one or more multicore processors that share a
large memory bank.  The algorithm developed in this paper targets a general
machine with this layout and the results are highlighted on Blue Waters, a Cray
machine at the National Center for Supercomputing Applications.  Blue Waters
consists of $22\,640$ Cray XE nodes, each containing two AMD 6276 Interlagos
processors for a total of 16 cores per node, and $4\,228$ Cray XK nodes
consisting of a single AMD processor along with an NVIDIA Kepler 
GPU\footnote{\url{https://bluewaters.ncsa.illinois.edu/hardware-summary}}.  The
nodes are connected through a three-dimensional torus Gemini interconnect,
with each Gemini serving two nodes.  The remainder of this paper with focus on
only the Cray XE nodes within Blue Waters.

Consider a system with $n_{p}$ processes distributed across $n_{n}$
nodes, resulting in $\ppn$ processes per node.  Rank $r \in \left[0,
n_{p}-1\right]$ is described by the tuple $(p, n)$ where $0 \leq p < \ppn$
is the local process number of rank $r$ on node $n$.  Assuming SMP-style ordering,
the first $\ppn$ ranks are mapped to the
first node, the next $\ppn$ to the second node, and so on.  Therefore,
rank $r$ is described by the tuple $\left( r \mod \ppn, \lfloor
\frac{r}{\ppn} \rfloor \right)$.  Thus, for the remainder of the paper, the notation of
rank $r$ is interchangeable with $\left(p, n\right)$.

Parallel matrices and vectors are distributed across all $n_{p}$ ranks such
that each process holds a portion of the linear system.  Let $\mathcal{R}(r)$
be the rows of an $N \times N$ sparse linear system, $w \leftarrow A \cdot v$,
stored on rank $r$.  In the case of an even, contiguous partition
where the $k^{\text{th}}$ partition is placed on the $k^{th}$ rank,
$\mathcal{R}(r)$ is defined as
\begin{align}
    \mathcal{R}(r) &=
    \left\{\left\lfloor\frac{N}{n_{p}}\right\rfloor r , \ldots,
    \left\lfloor\frac{N}{n_{p}}\right\rfloor (r+1) - 1\right\} \label{eq:R}\\
  \intertext{or equivalently as}
    \mathcal{R}((p, n)) &=
    \left\{\left\lfloor\frac{N}{n_{p}}\right\rfloor(p, n), \ldots,
    \left\lfloor\frac{N}{n_{p}}\right\rfloor ((p, n) +1) - 1\right\}
    \label{eq:R_pn}.
\end{align}
The rows of a matrix $A$ are partitioned into on-process and off-process blocks,
as described in Section~\ref{section:intro}.  Accounting for parallel nodal
awareness, the off-process block is further partitioned into on-node and
off-node blocks, as described in Example~\ref{example:simple_matrix}.
\begin{exmp}\label{example:simple_matrix}
    Suppose the parallel system consists of six processes distributed across
    three nodes, as displayed in Figure~\ref{figure:example_system}.
        \begin{figure}[!ht]
        \centering \includegraphics[width=.4\textwidth]{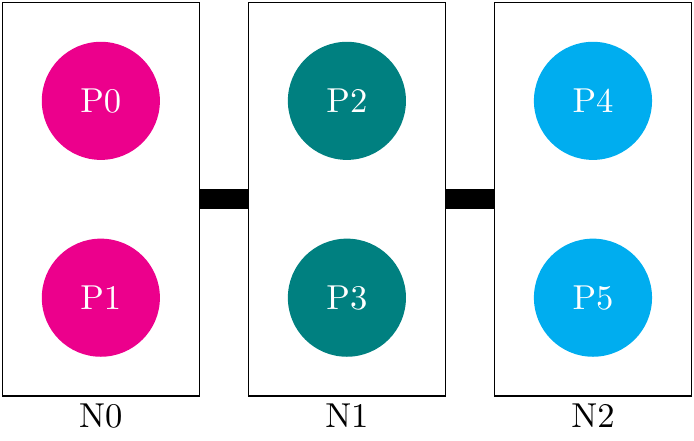}
        \caption{An example parallel system with six processes distributed
        across three nodes.}\label{figure:example_system}
    \end{figure}
        Let the linear
    system $w \leftarrow A \cdot v$ displayed in
    Figure~\ref{figure:example_matrix} be partitioned across this processor layout
    with each process holding a single row of the matrix and associated row of
    the input vector.
        \begin{figure}[!ht]
        \centering \includegraphics[width=.6\textwidth]{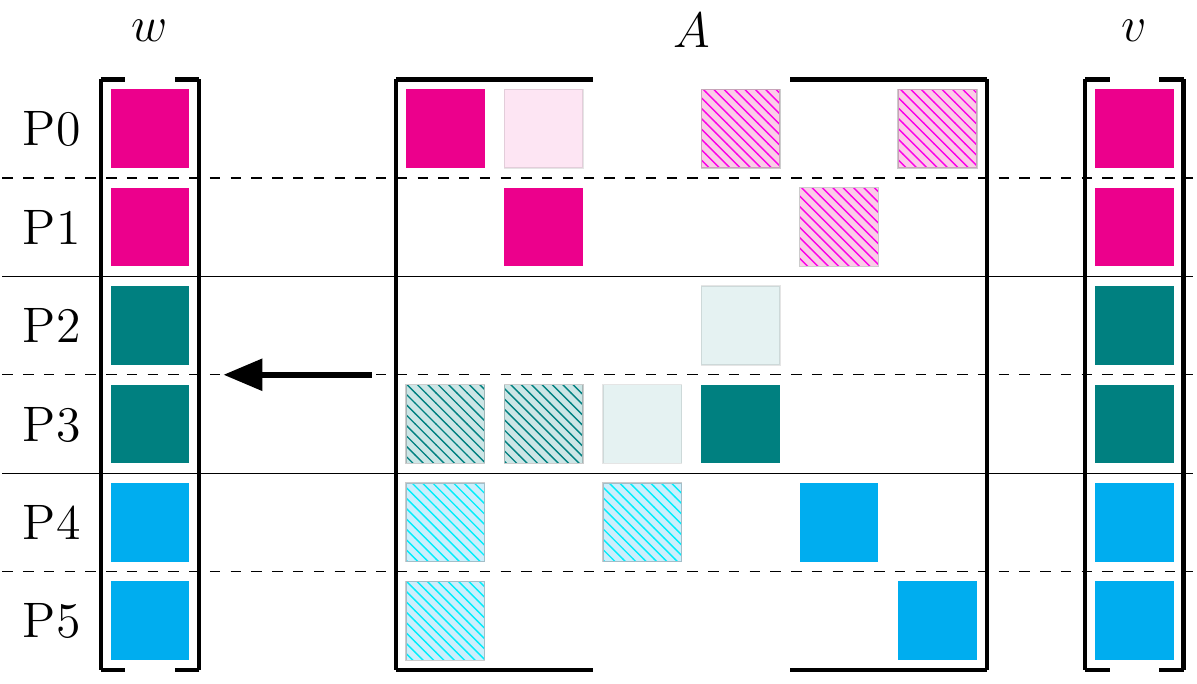}
        \caption{An example $6 \times 6$ sparse matrix for the parallel system in
        Figure~\ref{figure:example_system}.  The solid shading denotes blocks that
        require only on-node communication, while the striped shading denotes blocks
        that require communication with distant nodes.}\label{figure:example_matrix}
    \end{figure}
        In this example, the diagonal entry falls into the
    on-process block, as the corresponding vector value is stored locally.  The
    off-process block, which requires communication, consists of all
    off-diagonal non-zeros as the associated vector values are stored on other
    processes.
\end{exmp}
For any process $(p, n)$, the on-node columns of $A$ correspond to vector values
that are stored on some process $(s, n)$, where $s \neq p$.  Similarly, the
off-node columns of $A$ correspond to vector values stored on some process $(q, m)$,
where $m \neq n$.  To make this clearer, we define the following
\begin{align}
    \texttt{on\_process}(A, (p, n)) &=
    \left\{ A_{ij} \neq 0 \,\middle|\, i, j \in \mathcal{R}((p, n))
    \right\} \label{eq:on_proc}
\end{align}
\begin{multline}
    \texttt{off\_process}(A, (p, n)) =\\
    \left\{ A_{ij} \neq 0 \,\middle|\, i \in \mathcal{R}((p, n)), j
    \not\in
    \mathcal{R}((p, n)) \right\} \label{eq:off_proc}
\end{multline}
\begin{multline}
    \texttt{on\_node}(A, (p, n)) = \\
    \left\{ A_{ij} \neq 0 \,\middle|\, \exists
    \, q \neq p\, \textnormal{with $i \in\ \mathcal{R}((p, n))$, $j \in\
    \mathcal{R}((q, n))$} \right\} \label{eq:on_node}
\end{multline}
and
\begin{multline}
    \texttt{off\_node}(A, (p, n)) = \\
    \left\{ A_{ij} \neq 0 \,\middle|\, \exists
    \, q, m \neq n\, \textnormal{with  $i \in\ \mathcal{R}((p, n))$, $j \in\ \mathcal{R}((q, m))$} \right\}\label{eq:off_node}.
\end{multline}

\subsection{Standard SpMV}
For a sparse matrix-vector multiply, $w \leftarrow A \cdot v$, each process
receives all values of $v$ associated with the non-zero entries in the
off-process block of $A$.  For example, if rank $r$ contains a non-zero entry of
$A$, $A_{ij}$, at row $i$, column $j$, then rank $s$ with row $j \in
\mathcal{R}(s)$ sends the $j^{\text{th}}$ vector value, $v_{j}$, to rank $r$.
Typically, these communication requirements are determined as the sparse matrix
is formed~\cite{Bisseling2005,Schubert2011,Geus2001}.

In the reference SpMV, for each rank $r$ there is a list of processes to which
data is sent, as well as the global vector indices to be sent to each.  The
function $\mathcal{P}(r)$ defines the list of processes to which a rank $r$
sends.  Specifically,
\begin{equation}\label{eq:P}
  \mathcal{P}(r) = \left\{t \,\middle|\, \textnormal{$A_{ij}
\neq\ 0$ with $i\in\mathcal{R}{(t)}$, $j\in\mathcal{R}{(r)},\ r \neq t$}\right\}
\end{equation}
For each $t$ in $\mathcal{P}(r)$, define the function $\mathcal{D}(r, t)$ to
return the global vector indices that process $r$ sends to process $t$.  This
function is defined as follows.
\begin{equation}\label{eq:D}
  \mathcal{D}(r, t) = \left\{i \,\middle|\, \textnormal{$A_{ij}
\neq\ 0$ with $i\in\mathcal{R}{(t)}$, $j\in\mathcal{R}{(r)},\ r \neq t$}\right\}
\end{equation}

Consider a standard SpMV for the linear system described in
Example~\ref{example:simple_matrix}.  Table~\ref{table:example_P} lists the
processes to which each rank must send, while Table \ref{table:example_D}
displays the indices that each rank $r$ sends to any rank $t$.
\begin{table}[!ht]
   \centering
    \begin{tabular}{c c c c c c c}
    & \multicolumn{6}{c}{$r$}\\
    & 0 & 1 & 2 & 3 & 4 & 5 \\\cline{2-7}
    \multicolumn{1}{c|}{$\mathcal{P}(r)$} &
    $\left\{ 3, 4, 5 \right\}$ & 
    $\left\{ 0, 3 \right\}$ &
    $\left\{ 3, 4 \right\}$ &
    $\left\{ 0, 2 \right\}$ &    
    $\left\{ 1 \right\}$ &
    $\left\{ 0 \right\}$ \\
\end{tabular}

   \caption{Communication pattern for rank $r$ in
    Example~\ref{example:simple_matrix}, containing the values for
   $\mathcal{P}(r)$.}\label{table:example_P}.
\end{table}
\begin{table}[!ht]
    \centering
    \begin{tabular}{ c c c c c c c c }
                             & & \multicolumn{6}{ c }{$r$}\\
                             & & 0 & 1 & 2 & 3 & 4 & 5\\\cline{3-8}
    \multirow{6}{*}{$t$}     & \multicolumn{0}{c|}{0}
                             & $\left\{\right\}$ & $\left\{ 1 \right\}$
                             & $\left\{\right\}$ & $\left\{ 3 \right\}$
                             & $\left\{\right\}$ & $\left\{ 5 \right\}$ \\

                             & \multicolumn{1}{c|}{1}
                             & $\left\{\right\}$ & $\left\{\right\}$
                             & $\left\{\right\}$ & $\left\{\right\}$
                             & $\left\{ 4 \right\}$ & $\left\{\right\}$ \\

                             & \multicolumn{1}{c|}{2}
                             & $\left\{\right\}$ & $\left\{\right\}$
                             & $\left\{\right\}$ & $\left\{ 3 \right\}$
                             & $\left\{\right\}$ & $\left\{\right\}$ \\

                             & \multicolumn{1}{c|}{3}
                             & $\left\{ 0 \right\}$ & $\left\{ 1 \right\}$
                             & $\left\{ 2 \right\}$ & $\left\{\right\}$
                             & $\left\{\right\}$ & $\left\{\right\}$ \\

                             & \multicolumn{1}{c|}{4}
                             & $\left\{ 0 \right\}$ & $\left\{\right\}$
                             & $\left\{ 2 \right\}$ & $\left\{\right\}$
                             & $\left\{\right\}$ & $\left\{\right\}$ \\

                             & \multicolumn{1}{c|}{5}
                             & $\left\{ 0 \right\}$ & $\left\{\right\}$
                             & $\left\{\right\}$ & $\left\{\right\}$
                             & $\left\{\right\}$ & $\left\{\right\}$ \\
\end{tabular}

    \caption{      Each column $r$ lists the indices of values sent to each process $t$ in
      $\mathcal{P}(r)$, namely $\mathcal{D}(r,s)$.}\label{table:example_D}.
\end{table}

With these definitions, the \textit{standard} or reference SpMV is described in
Algorithm~\ref{alg:std_spmv}.
It is important to note that the parallel communication in
Algorithm~\ref{alg:std_spmv} is executed independent of any locality in the
problem.  That is, messages sent to another process may be both on-node or
off-node depending on the process, however this is not considered in the
algorithm.

\begin{algorithm2e}[!ht]
  \DontPrintSemicolon	\KwIn{    \begin{tabular}[t]{l l}
        $r$\\
        $A|_{\mathcal{R}(r)}$\\
        $v|_{\mathcal{R}(r)}$\\
    \end{tabular}
  }
  \BlankLine	\KwOut{    \begin{tabular}[t]{l l}
        $w|_{\mathcal{R}(r)}$
    \end{tabular}
  }
  \BlankLine  $A_{\text{on\_process}} = \code{on\_process}(A|_{\mathcal{R}(r)})$\;
  $A_{\text{off\_process}} = \code{off\_process}(A|_{\mathcal{R}(r)})$\;
    \For{$t \in \mathcal{P}(r)$}{        \For{$i \in \mathcal{D}(r, t)$} {          $b_{\text{send}} \leftarrow v|_{\mathcal{R}{(r)}_{i}}$\;
        }
        $\code{MPI\_Isend}(b_{\text{send}}, \ldots, t, \ldots)$\;
    }
    $b_{\text{recv}} \leftarrow \emptyset$\;
    \For{$t$ \textnormal{s.t.} $r \in \mathcal{P}(t)$}{        $\code{MPI\_Irecv}(b_{\text{recv}}, \ldots, t, \ldots)$\;
    }
    $\code{local\_spmv}(A_{\text{on\_process}}, v|_{\mathcal{R}(r)})$\;
    $\code{MPI\_Waitall}$\;
    $\code{local\_spmv}(A_{\text{off\_process}}, b_{\text{recv}})$\;
	\caption{\code{standard\_spmv}}\label{alg:std_spmv}
\end{algorithm2e}

\section{Communication Models}\label{section:models}

The performance of Algorithm~\ref{alg:std_spmv} is sub-optimal since it does
not take advantage of node locality in the communication.  To see this, a
communication performance model is developed in this section.  One approach
is that of the \textit{max-rate model}~\cite{2016GrOlSa_pingpong}, which
describes the communication time as
\begin{equation}
    T = \alpha + \frac{\ppn \cdot s}{\min(B_{\text{N}}, B_{\text{max}}
        + (\ppn - 1) B_{\text{inj}})},
    \label{eq:max_rate}
\end{equation}
where $\alpha$ is the \textit{latency} or start-up cost of a message, which may
include preparing a message for transport or determining the network route; $s$ is
the number of bytes to be communicated; \ppn\ is again the number of
communicating processes per node; $B_{\text{inj}}$ is the maximum
rate at which messages are injected into the network; $B_{\max}$ is the
achievable message rate of each process or
$\textit{bandwidth}$; and $B_{\text{N}}$ is the peak rate
of the network interface controller (NIC)\@.  In the simplest case of $\ppn=1$, the familiar
\textit{postal model} suffices:
\begin{equation}
    T = \alpha + \frac{s}{B_{\max}}.
\end{equation}

MPI contains multiple message passing protocols, including short, eager, and
rendezvous.  Each message consists of an envelope, including information about
the message such as message size and source information, as well as message
data.  Short messages contain very little data which is sent as part of the
envelope.  Eager and rendezvous messages, however, send the envelope followed by
packets of data.  Eager messages are sent under the assumption that the
receiving process has buffer space available to store data that is communicated.
Therefore, a message is sent without checking buffer space at the receiving
process, limiting the associated latency.  However, if a message is sufficiently
large, rendezvous protocol must be used.  This protocol requires the sending
process to inform the receiving rank of the message so that buffer space is
allocated.  The message is sent only once the sending process is informed that
this space is available.  Therefore, there is a larger overhead with sending a
message using rendezvous protocol.  Table~\ref{table:inter_comm} displays the
measurements for $\alpha$, $B_{\text{inj}}$, $B_{\max}$, and
$B_{\text{N}}$ for Blue Waters, as determined for the \textit{max-rate}
model.
\begin{table}
    \centering
    \begin{tabular} { c c c c c }
  \toprule
    & $\alpha$ & $B_{\text{inj}}$ & $B_{\max}$ & $B_{\text{N}}$ \\
  \midrule
    Short & $4.0 \cdot 10^{-6}$ & $6.3 \cdot 10^{8}$ & $-1.8 \cdot 10^{7}$ &
    $\infty$\\
    Eager & $1.1 \cdot 10^{-5}$ & $1.7 \cdot 10^{9}$ & $6.2 \cdot 10^{7}$ &
    $\infty$\\
    Rend & $2.0 \cdot 10^{-5}$ & $3.6 \cdot 10^{9}$ & $6.1 \cdot 10^{8}$
    & $5.5 \cdot 10^{9}$\\
    \bottomrule
\end{tabular}

    \caption{Measurements for $\alpha$, $B_{\text{inj}}$, $B_{\min}$,
    and $B_{\text{N}}$ for Blue Waters.}\label{table:inter_comm}
\end{table}

The \textit{max-rate} model can be improved by distinguishing between intra- and
inter-node communication.  If the sending and receiving processes lie on the
same physical node, data is not injected into the network, yielding low start-up
and byte transport costs.  As intra-node messages are not injected into the
network, communication local to a node can be modeled as
\begin{equation}
    T_{\ell} = \alpha_{\ell} + \frac{s_{\ell}}{B_{\max_{\ell}}},
    \label{eq:intra_comm}
\end{equation}
where $\alpha_{\ell}$ is the start-up cost for intra-node messages;
$s_{\ell}$ is the number of bytes to be transported; and $B_{\max_{\ell}}$
is the achievable intra-node message rate.

Nodecomm\footnote{See \url{https://bitbucket.org/william_gropp/baseenv}}, 
a topology-aware communication program, measures the time
required to communicate on various levels of the parallel system, such as
between two nodes of varying distances and between processes local to a node.
Communication tests between processes local to one node were used to calculate
the intra-node model parameters, as displayed in Table~\ref{table:intra_comm}.
\begin{table}
    \centering
    \begin{tabular} { c c c c }
  \toprule
    & $\alpha_{\ell}$ & $B_{\max_{\ell}}$ \\
  \midrule
    Short & $1.3 \cdot 10^{-6}$ & $4.2 \cdot 10^{8}$ \\
    Eager & $1.6 \cdot 10^{-6} $ & $7.4 \cdot 10^{8}$\\
    Rend & $4.2 \cdot 10^{-6}$ & $3.1 \cdot 10^{9}$\\
    \bottomrule
\end{tabular}

    \caption{Measurements for intra-node variables, $\alpha_{\ell}$ and
    $B_{\max_{\ell}}$.}\label{table:intra_comm}
\end{table}

\begin{figure}[!ht]
    \centering
    \includegraphics[width=0.6\textwidth]{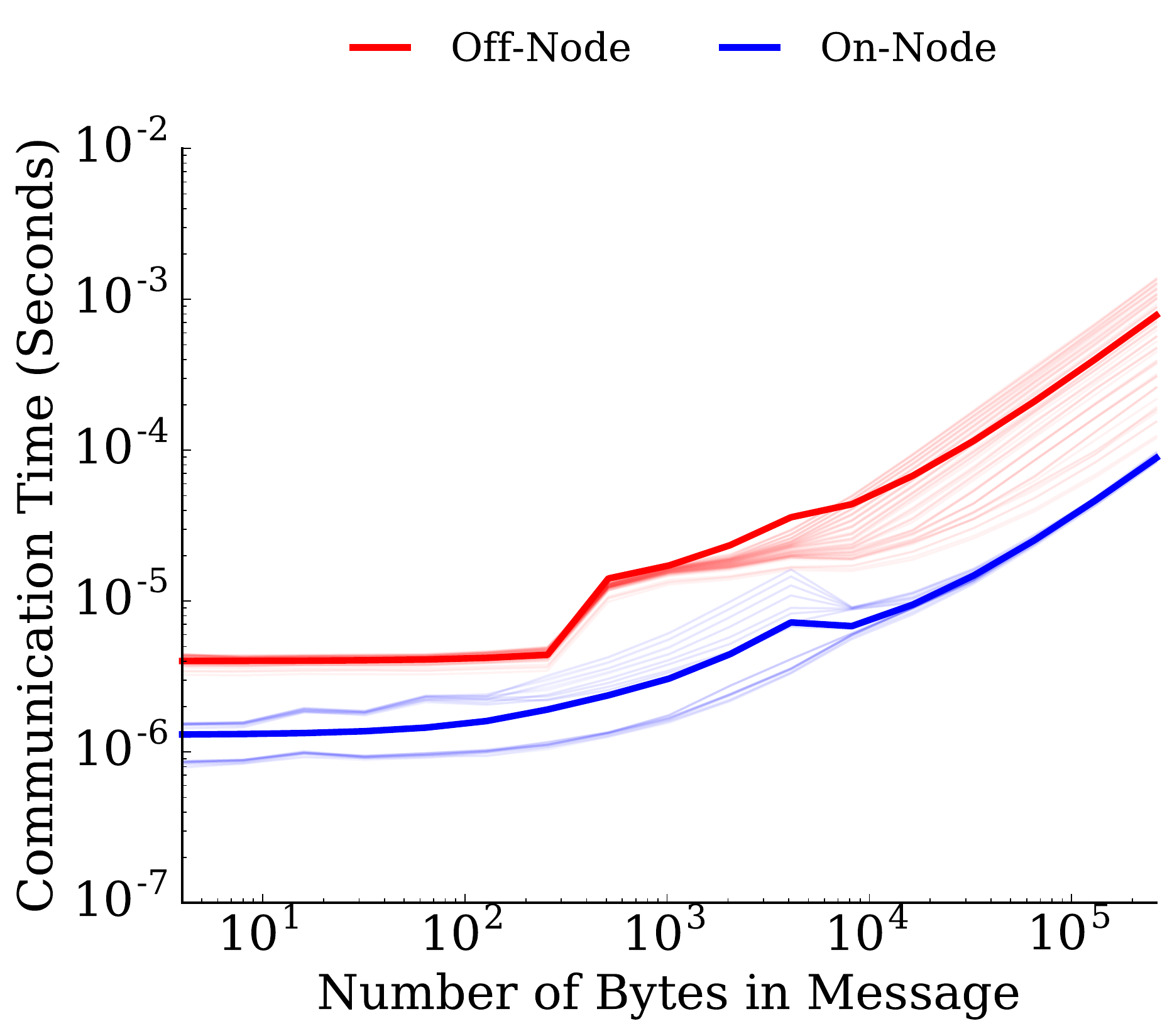}
    \caption{The time required to send a single message of various sizes,
    with the thin lines representing timings measured by Nodecomm and the thick
    lines displaying the \textit{max-rate} and intra-node models in~\eqref{eq:max_rate}~and~\eqref{eq:intra_comm}, respectively.}\label{figure:model_msg_times}
\end{figure}

Furthermore, Figure~\ref{figure:model_msg_times} shows the time required to
send a single message of varying sizes.  The thin lines display Nodecomm
measurements for time required to send a single message, as either inter- or
intra-node communication.  Furthermore, the thick lines represent the time
required to send a message of each size, according to the \textit{max-rate}
model in~\eqref{eq:max_rate} and intra-node model in~\eqref{eq:intra_comm}.  This figure
displays a significant difference between the costs of intra- and inter-node
communication.

\section{Node Aware Parallel SpMV}\label{section:tapspmv}

To reduce communication costs, the algorithm proposed in this section decreases
the number and size of messages being injected into the network by increasing
the amount of intra-node communication, which is less-costly than inter-node
communication.  This trade-off is accomplished through a so-called
\textit{node aware parallel} SpMV (NAPSpMV), where values are gathered in processes
local to each node before being sent across the network, followed by a
distribution of processes on the receiving node.  As a result, as the matrix is
formed each process $(p, n)$ determines the communicating processes during the
various steps of a NAPSpMV, as well as the accompanying data.  A high level
overview of the process is described in Example~\ref{example:H_overview}.  It
is important to note that the communication for each NAPSpMV is load-balanced
such that all processes local to node $n$ send and receive both a similar number
and size of messages through inter-node communication.  Therefore, it is assumed
that the nodes $n$ and $m$ in Example~\ref{example:H_overview} are only a
portion of the parallel system, and $n$ is communicating with other nodes in a
similar fashion.  If the parallel system consists only of nodes $n$ and $m$,
each process on node $n$ would send a portion of the data to node $m$.
\begin{exmp}\label{example:H_overview}
    During each NAPSpMV, off-node data is communicated through a three-step
    process, as displayed in Figure~\ref{figure:H_overview}.
    \begin{figure}
        \centering
        \includegraphics[width=0.6\textwidth]{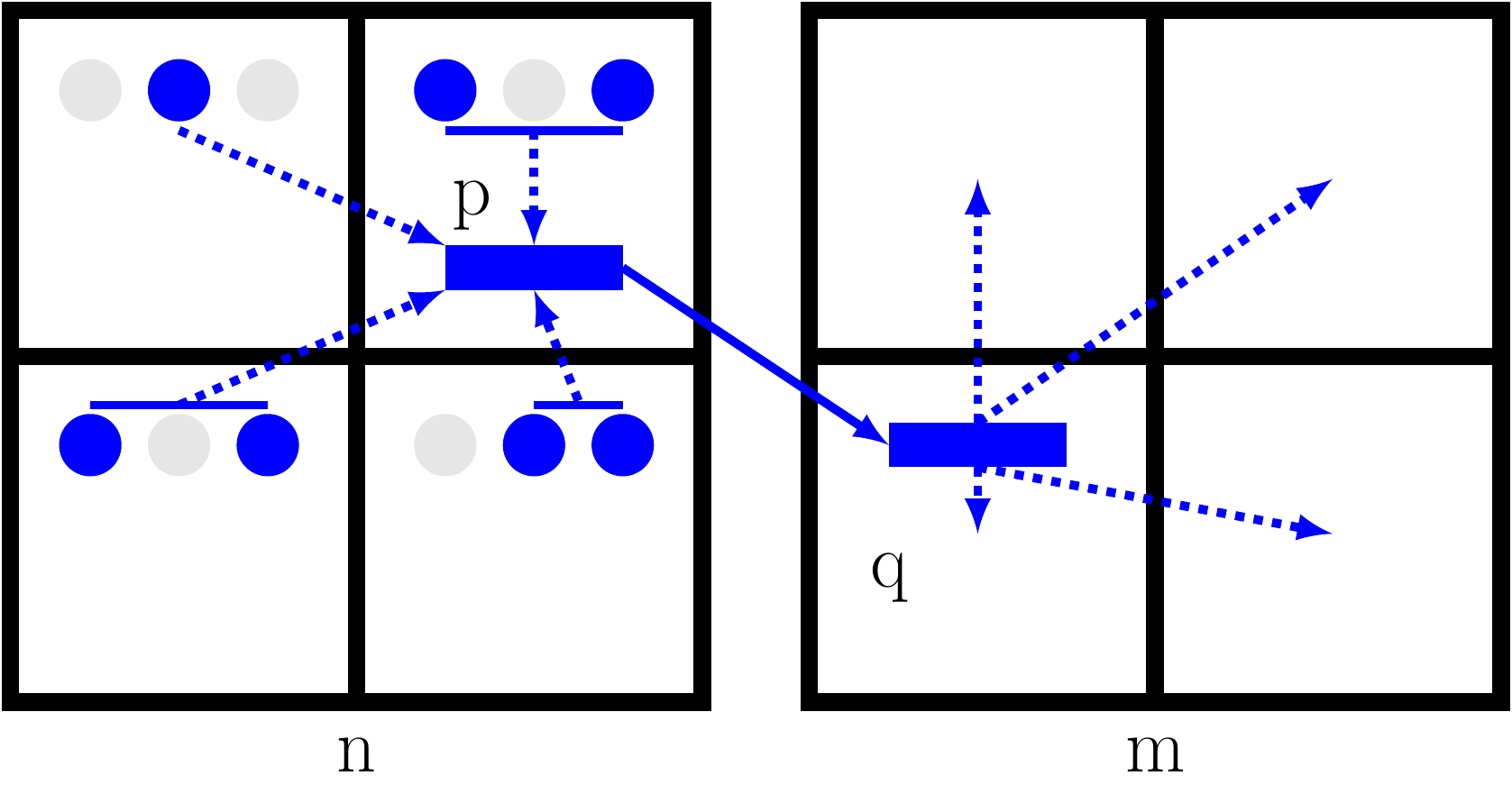}
        \caption{The various arrows exemplify the process of communicating data
        from each process on node $n$ to processes on node $m$ through a three-step
        algorithm.  The bold circles on node $n$ represent vector values that must be
        communicated to node $m$.}\label{figure:H_overview}
    \end{figure}
        This figure displays a portion of a parallel system consisting of 8
    processes partitioned across two nodes, labeled $n$ and $m$.  The solid
    circles on node $n$ represent vector values that must be sent to node $m$.
    Therefore, each process on node $n$ must send values to processes on node
    $m$.  Instead of sending directly to destination processes, each process
    $(s, n)$ sends to the process labeled $(p, n)$, displayed by the dashed
    arrows on node $n$.  Process $(p, n)$ then sends all collected values
    through the network to process $(q, m)$.  Finally, process $(q, m)$
    distributes received values among the processes local to node $m$, displayed
    by the dashed arrows on node $m$.

    On-node data is communicated directly between the process on which the
    vector values are stored and that which requires the data, as displayed in
    Figure~\ref{figure:H_overview_local}.
        \begin{figure}
        \centering
        \includegraphics[width=0.3\textwidth]{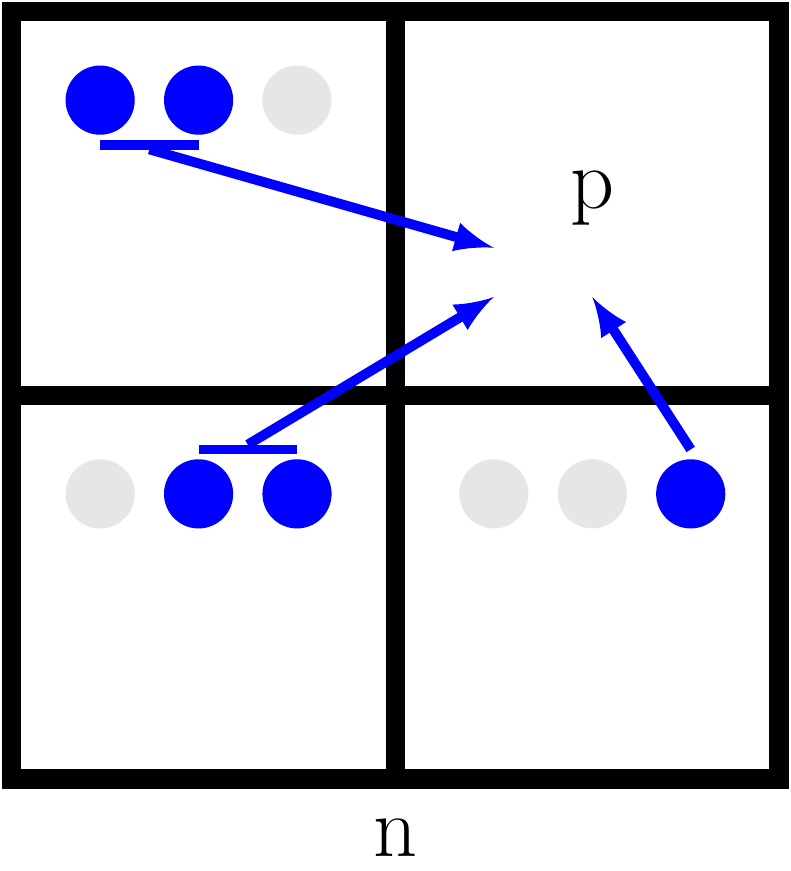}
        \caption{An example of how vector values corresponding with matrix
        entries in the on-node block are communicated.  All values $(p, n)$ must
        receive from other processes $(q, n)$ are communicated directly as
        nothing is injected into the network.}\label{figure:H_overview_local}
    \end{figure}
        In this example, the solid circles represent vector values that are stored
    on each process $(s, n)$ and needed by $(p, n)$.  This data is sent directly
    between the processes in a single step.
\end{exmp}

\subsection{Inter-node communication setup}

To eliminate the communication of duplicated messages, a list of communicating
nodes is formed for each node $n$ along with the accompanying data values.
These lists are then distributed across all processes local to $n$ by balancing
the number of nodes and volume of data for communication.  To facilitate this,
the function $\mathcal{N}(n)$ defines the set of nodes to which the processes on
node $n$ must send,
\begin{equation}
    \begin{split}
    \mathcal{N}(n) &= \left\{m \,\middle|\, \textnormal{$\exists$ $p$, $q$ s.t.
        $A_{ij} \neq\ 0$} \right.\\ & \left. \textnormal{with $i \in\
        \mathcal{R}{((q, m))}$, $j \in\ \mathcal{R}{((p, n))}$, $n \neq\ m$}
        \vphantom{m} \right\}. \label{eq:N}
    \end{split}
\end{equation}

Table~\ref{table:example_N} contains $\mathcal{N}(n)$, the list of nodes to
which each node $n$ sends.
\begin{table}[!ht]
    \centering
    \begin{tabular}{c c c c}
                                            & \multicolumn{3}{c}{$n$}\\
                                            & 0 & 1 & 2 \\\cline{2-4}
    \multicolumn{1}{c|}{$\mathcal{N}(n)$}  & 
    $\left\{ 1, 2 \right\}$ & 
    $\left\{ 0, 2 \right\}$ &
    $\left\{ 0 \right\}$ \\
\end{tabular}

    \caption{Communication requirements for each node $n$ in
    Example~\ref{example:simple_matrix}.}\label{table:example_N}
\end{table}
The associated data values are defined for each node $m \in
\mathcal{N}(n)$ with $\mathcal{E}(n, m)$, which returns the data
indices to be sent from node $n$ to node $m$.  That is,
\begin{multline}\label{eq:E}
  \mathcal{E}(n, m) = \left\{ i \,|\, \exists\ \textnormal{$p$, $q$ s.t. $A_{ij}
    \neq\ 0$ with $i \in \mathcal{R}{((q, m))}$}\right.,\\
                                      \left.   \textnormal{$j \in
                                      \mathcal{R}{((p, n))}$, $n \neq\ m$}\right\}.
\end{multline}
Extending Example~\ref{example:simple_matrix},
Table~\ref{table:example_E} displays the global vector indices,
$\mathcal{E}(n, m)$, for each set of nodes $n$ and $m$.
\begin{table}[!ht]
    \centering
    \begin{tabular}{ c c c c c }
                             & & \multicolumn{3}{ c }{$n$}\\
                             & & 0 & 1 & 2\\\cline{3-5}
        \multirow{3}{*}{$m$} 
        & \multicolumn{1}{c|}{0} 
        & $\left\{\right\}$ 
        & $\left\{ 3 \right\}$ 
        & $\left\{ 4, 5 \right\}$ \\

        & \multicolumn{1}{c|}{1} 
        & $\left\{ 0, 1 \right\}$ 
        & $\left\{  \right\}$ 
        & $\left\{  \right\}$ \\

        & \multicolumn{1}{c|}{2} 
        & $\left\{ 0 \right\}$ 
        & $\left\{ 2 \right\}$ 
        & $\left\{\right\}$ \\
\end{tabular}

    \caption{In Example~\ref{example:simple_matrix},
    each column $n$ contains the values sent from $n$ to $m$, as in $\mathcal{E}(n,
    m)$.}\label{table:example_E}
\end{table}

$\mathcal{T}((p, n))$ defines the nodes to which $(p, n)$ must send, that is the
nodes in $\mathcal{N}(n)$ that are distributed to process $(p, n)$.  Similarly,
$\mathcal{U}((p, n))$ contains the nodes that send to $(p, n)$.  Specifically,
\begin{align}
    \mathcal{T}((p, n)) &= \left\{m \in
    \mathcal{N}(n) \,\middle|\, m
    \;\textnormal{maps to}\; (p,n) \right\}, \label{eq:T}\\
    \mathcal{U}((p, n)) &= \left\{m \,\middle|\, n \in
    \mathcal{N}(m),\ n
    \;\textnormal{maps to}\; (p, n) \right\}. \label{eq:U}
\end{align}

This paper considers a simple distribution in which the node $m \in
\mathcal{N}(n)$ to which the most data $\left|\mathcal{D}(n, m)\right|$ is 
sent is mapped to process $(0, n)$, the node with the second most data is mapped
to process $(1, n)$, and so on.  The opposite ordering is used for
$\mathcal{U}((p, n))$, mapping the node $n \in \mathcal{N}(m)$ with largest 
$\left|\mathcal{D}(m, n)\right|$ to process $(\ppn - 1, n)$, the second largest
to process $(\ppn - 2, n)$, etc.  If there are fewer nodes in $\mathcal{N}(n)$
than there are processes per node, a single node is mapped to multiple local
processes so that all processes communicate.  There are various other possible
mapping strategies, such as mapping a node $m$ to the process $(p, n)$ storing the
majority of the data in $\mathcal{D}(n, m)$.  However, as this would only affect
intra-node communication requirements, these mappings are not explored in this
paper.

The processor layout in Example~\ref{example:simple_matrix} is
displayed in Table~\ref{table:example_T}, where the columns contain the send
and receive nodes that are mapped to each process.
\begin{table}[!ht]
    \centering
    \begin{tabular}{ c c c c c c c }
                             & \multicolumn{6}{ c }{$(p, n)$}\\
                             & (0, 0) & (1, 0) & (0, 1) & (1, 1) & (0, 2) & (1,
                             2)\\\cline{2-7} 
       \multicolumn{1}{c|}{
           $\mathcal{T}((p, n))$
      }
        & $\left\{1\right\}$
        & $\left\{2\right\}$
        & $\left\{0\right\}$
        & $\left\{2\right\}$
        & $\left\{0\right\}$
        & $\left\{\right\}$\\
       \multicolumn{1}{c|}{
           $\mathcal{U}((p, n))$
      }
        & $\left\{2\right\}$
        & $\left\{1\right\}$
        & $\left\{\right\}$
        & $\left\{0\right\}$
        & $\left\{1\right\}$
        & $\left\{0\right\}$\\
\end{tabular}

    \caption{Processor mappings for $\mathcal{N}(n)$, namely $\mathcal{T}((p, n))$ and
    $\mathcal{U}((p, n))$ for
    Example~\ref{example:simple_matrix}.}\label{table:example_T}
\end{table}

Finally, $\mathcal{G}((p, n))$ defines the set of all
off-node processes to which process $(p, n)$ sends data during the inter-node
communication step of the NAPSpMV\@.  Specifically,
\begin{align}
    \mathcal{G}((p, n)) &= \left\{(q, m) \,\middle|\, m \in
    \mathcal{T}((p, n)),\, n \in \mathcal{U}((q, m)) \right\}. \label{eq:G}
\end{align}
Following Example~\ref{example:simple_matrix}, the columns of
Table~\ref{table:example_G} list the indices of the values that each $(p, n)$
sends,
$\mathcal{G}((p, n))$.
\begin{table}[!ht]
    \centering
    \begin{tabular}{ c c c c c c c }
                             & \multicolumn{6}{ c }{$(p, n)$}\\
                             \multirow{1}{*}{
      }
                             & (0, 0) & (1, 0) & (0, 1) & (1, 1) & (0, 2) & (1,
                             2)\\\cline{2-7}

           \multicolumn{1}{c|}{$\mathcal{G}((p, n))$}
        & $\left\{(1, 1)\right\}$
        & $\left\{(1, 2)\right\}$
        & $\left\{(1, 0)\right\}$
        & $\left\{(0, 2)\right\}$
        & $\left\{(0, 0)\right\}$
        & $\left\{\right\}$\\

\end{tabular}

    \caption{Inter-node communication requirements of each process $(p, n)$ for
    Example~\ref{example:simple_matrix}}.\label{table:example_G}
\end{table}
Finally, let $\mathcal{I}((p, n), (q, m))$ define the global data indices
corresponding to the values sent from process $(p, n)$ to $(q, m)$:
\begin{multline}
    \mathcal{I}((p, n), (q, m)) =\\
    \left\{\mathcal{E}(n, m)
    \,\middle|\,  m \in
    \mathcal{T}((p, n)),\, n \in \mathcal{U}((q, m)
    \right\}\label{eq:IGset_send}
\end{multline}
The global vector indices to which each process $(p, n)$ sends and receives
for Example~\ref{example:simple_matrix} are displayed in
Table~\ref{table:example_I}.
\begin{table}[!ht]
    \centering
    \begin{tabular}{ c c c c c c c c }
                             & \multicolumn{6}{ c }{$(p, n)$}\\
                             & & (0, 0) & (1, 0) & (0, 1) & (1, 1) & (0, 2) & (1,
  2)\\\cline{3-8} 
        \multirow{6}{*}{\begin{sideways}$(q, m)$\end{sideways}}  
        & \multicolumn{1}{c|}{(0, 0)} 
                       & $\left\{  \right\}$ 
                       & $\left\{  \right\}$  
                       & $\left\{  \right\}$  
                       & $\left\{  \right\}$  
                       & $\left\{ 4, 5 \right\}$  
                       & $\left\{  \right\}$ \\
        & \multicolumn{1}{c|}{(1, 0)}  
                       & $\left\{  \right\}$ 
                       & $\left\{  \right\}$  
                       & $\left\{ 3 \right\}$  
                       & $\left\{  \right\}$  
                       & $\left\{  \right\}$  
                       & $\left\{  \right\}$ \\
        & \multicolumn{1}{c|}{(0, 1)}                        
                       & $\left\{  \right\}$ 
                       & $\left\{  \right\}$  
                       & $\left\{  \right\}$  
                       & $\left\{  \right\}$  
                       & $\left\{  \right\}$  
                       & $\left\{  \right\}$ \\
        & \multicolumn{1}{c|}{(1, 1)} 
                       & $\left\{ 0 \right\}$ 
                       & $\left\{  \right\}$  
                       & $\left\{  \right\}$  
                       & $\left\{  \right\}$  
                       & $\left\{  \right\}$  
                       & $\left\{  \right\}$ \\
        & \multicolumn{1}{c|}{(0, 2)} 
                       & $\left\{  \right\}$ 
                       & $\left\{  \right\}$  
                       & $\left\{  \right\}$  
                       & $\left\{ 2 \right\}$  
                       & $\left\{  \right\}$  
                       & $\left\{  \right\}$ \\
        & \multicolumn{1}{c|}{(1, 2)} 
                       & $\left\{  \right\}$ 
                       & $\left\{ 0, 1 \right\}$  
                       & $\left\{  \right\}$  
                       & $\left\{  \right\}$  
                       & $\left\{  \right\}$  
                       & $\left\{  \right\}$ \\
\end{tabular}

    \caption{Inter-node communication requirements for each set of processes $(p, n)$
    and $(q, m)$.  Each column $(p, n)$ contains the indices of values sent from $(p, n)$
    to $(q, m)$.}\label{table:example_I}
\end{table}

\subsection{Local Communication} \label{section:local}

The function $\mathcal{G}_{\text{send}}((p, n))$ for $p = 0, \ldots,
\ppn - 1$, describes evenly distributed inter-node communication requirements
for all processes local to node $n$.  However, many of the vector indices to be
sent to off-node process $(q, m) \in \mathcal{D}((p, n), (q, m))$, are not stored on process $(p, n)$.  For instance, in
Table~\ref{table:example_I}, process $(0, 1)$ sends global vector indices
$0$ and $1$.  However, only row $1$ is stored on process $(0, 1)$, requiring
vector component
$0$ to be communicated before inter-node messages are sent.

Similarly, many of the indices that a process $(q, m)$
receives from $(p, n)$ are redistributed to various processes on node $n$.
Table~\ref{table:example_I} requires process $(1, 2)$ to receive vector data
according to indices $0$ and $1$.  Process $(0, 2)$ uses both of these vector
values, yielding a requirement for redistribution of data received from
inter-node communication.  Therefore, local communication requirements must be
defined.

Each NAPSpMV consists of multiple steps of intra-node communication.  Let a
function $\mathcal{L}((p, n), \locality)$ define all processes, local to node
$n$, to which process $(p, n)$ sends messages, where $\locality$ is a tuple
describing the locality of both the original location of the data as well as its
final destination.  The locality of each position is described as either
$\texttt{on\_node}$, meaning a process local to node $n$, or
$\texttt{off\_node}$, meaning a process local to node $m \neq n$.

There are three possible combinations for $\locality$: 1.\ the data is
initialized $\texttt{on\_node}$ with a final destination $\texttt{off\_node}$;
2.\ the original data is $\texttt{off\_node}$ while the final destination is
$\texttt{on\_node}$; or 3.\ both the original data and the final location are
$\texttt{on\_node}$.  These three types of intra-node communication are
described in more detail in the remainder of Section~\ref{section:local}.

For each process $(s, n) \in \mathcal{L}((p, n), \locality)$,
$\mathcal{J}((p, n), (s, n), \locality)$ defines the
global vector indices to be sent from process $(p, n)$ to $(s, n)$
through intra-node communication.  This notation is used in following sections.

\subsubsection{Local redistribution of initial data}\label{section:L0}~\\

During inter-node communication, a process $(p, n)$ sends all vector values
corresponding to the global indices in $\mathcal{I}((p, n), (q, m))$ to each
process $(q, m) \in \mathcal{G}((p, n))$.  The indices in $\mathcal{I}((p, n),
(q, m))$ originate on node $n$, but not necessarily process $(p, n)$.
Therefore, the initial vector values must be redistributed among all processes
local to node $n$.

Let $\mathcal{L}((p, n), (\texttt{on\_node}, \texttt{off\_node}))$ represent all
processes, local to node $n$, to which $(p, n)$ sends initial vector values.
This function is defined as
\begin{multline}
    \mathcal{L}((p, n), (\texttt{on\_node}, \texttt{off\_node})) =\\ 
    \left\{(s, n) \,\middle|\, \exists \, j \in \mathcal{R}((p, n)),\ j \in
    \mathcal{I}((s, n), (q, m))\right\}. \label{eq:L0}
\end{multline}
The local processes to which each $(p, n)$ sends initial data in
Example~\ref{example:simple_matrix} are displayed in
Table~\ref{table:example_L0}.
\begin{table}[!ht]
    \centering
    \begin{tabular}{ c c c c c c c }
                             & \multicolumn{6}{ c }{$(p, n)$}\\
                             & (0, 0) & (1, 0) & (0, 1) & (1, 1) & (0, 2) & (1,
                             2)\\\cline{2-7}
        \multicolumn{1}{c|}{$\mathcal{L}$}
        & $\left\{(1, 0)\right\}$
        & $\left\{\right\}$
        & $\left\{(1, 1)\right\}$
        & $\left\{(0, 1)\right\}$
        & $\left\{\right\}$
        & $\left\{(0, 2)\right\}$\\

\end{tabular}

    \caption{Initial intra-node communication requirements for each process $(p,
    n)$ in Example~\ref{example:simple_matrix}.  The row of the
    table describes $\mathcal{L}((p, n),
    (\texttt{on\_node}, \texttt{off\_node}))$.}\label{table:example_L0}
\end{table}

Furthermore, the data global vector indices that must be sent from process $(p,
n)$ to each $(s, n) \in \mathcal{L}((p, n),
(\texttt{on\_node},\texttt{off\_node}))$ are defined as
\begin{multline}
    \mathcal{J}((p, n), (s, n), (\texttt{on\_node},
    \texttt{off\_node})) =\\
    \left\{i \,\middle|\, i \in \mathcal{R}((p, n)), \forall
    \, i \in
    \mathcal{G}((s, n))\right\}.\label{eq:IL0}
\end{multline}
The global vector indices that each $(p, n)$ must send to other processes on
node $n$ in Example~\ref{example:simple_matrix} are displayed in
Figure~\ref{table:example_J0}.
\begin{table}[!ht]
    \centering
    \begin{tabular}{ c c c c c c c c }
       & \multicolumn{6}{ c }{$(p, n)$}\\
       & & (0, 0) & (1, 0) & (0, 1) & (1, 1) & (0, 2) & (1, 2)\\\cline{3-8}
        \multirow{6}{*}{\begin{sideways}$(q, n)$\end{sideways}}
        & \multicolumn{1}{c|}{(0, 0)}
                       & $\left\{  \right\}$
                       & $\left\{  \right\}$
                       & ---
                       & ---
                       & ---
                       & --- \\
        & \multicolumn{1}{c|}{(1, 0)}
                       & $\left\{ 0 \right\}$
                       & $\left\{  \right\}$
                       & ---
                       & ---
                       & ---
                       & --- \\
        & \multicolumn{1}{c|}{(0, 1)}
                       & ---
                       & ---
                       & $\left\{  \right\}$
                       & $\left\{ 3 \right\}$
                       & ---
                       & --- \\
        & \multicolumn{1}{c|}{(1, 1)}
                       & ---
                       & ---
                       & $\left\{ 2 \right\}$
                       & $\left\{  \right\}$
                       & ---
                       & --- \\
        & \multicolumn{1}{c|}{(0, 2)}
                       & ---
                       & ---
                       & ---
                       & ---
                       & $\left\{  \right\}$
                       & $\left\{ 5 \right\}$ \\
        & \multicolumn{1}{c|}{(1, 2)}
                       & ---
                       & ---
                       & ---
                       & ---
                       & $\left\{  \right\}$
                       & $\left\{  \right\}$ \\
\end{tabular}

    \caption{Global vector indices of initial data that is communicated between
    processes local to each node $n$ in Example~\ref{example:simple_matrix}.
    Each column contains the indices of values sent
    from $(p, n)$ to $(q, n)$.  Note: dashes (---) throughout the table
    represent processes on separate nodes, which do not communicate during
    intra-node communication.}\label{table:example_J0}
\end{table}

\subsubsection{Local redistribution of received off-node data}\label{section:L1}~\\

During inter-node communication, a process $(p, n)$ sends all data with final
destination on node $m$ to process $(q, m) \in \mathcal{G}((p, n))$.  Process
$(q, m)$ then distributes these values across the processes local to node $m$.
Let $\mathcal{L}((q, m), (\texttt{off\_node}, \texttt{on\_node}))$ define all
processes local to node $m$ to which process
$(q, m)$ sends vector values that have been received through inter-node
communication.  This function is defined as
\begin{multline}
    \mathcal{L}((q, m), (\texttt{off\_node}, \texttt{on\_node})) =\\ 
    \left\{(s, m) \,\middle|\, \exists
    \, A_{ij} \neq 0 \;\textnormal{with}\; i \in \mathcal{R}((s, m)),\right.\\
    \left. j \in
    \mathcal{I}((p, n), (q, m))\right\}. \label{eq:L1_send}
\end{multline}
This is highlighted, for Example~\ref{example:simple_matrix}, in
Table~\ref{table:example_L1}.
\begin{table}[!ht]
    \centering
    \begin{tabular}{ c c c c c c c }
                             & \multicolumn{6}{ c }{$(p, n)$}\\
                             & (0, 0) & (1, 0) & (0, 1) & (1, 1) & (0, 2) & (1,
                             2)\\\cline{2-7}
        \multicolumn{1}{c|}{$\mathcal{L}$}
        & $\left\{\right\}$
        & $\left\{(0, 0)\right\}$
        & $\left\{\right\}$
        & $\left\{\right\}$
        & $\left\{\right\}$
        & $\left\{(0, 2)\right\}$\\
\end{tabular}

    \caption{Intra-node communication requirements containing processes to which
    each $(p, n)$ sends received inter-node data, according to
    Example~\ref{example:simple_matrix}.  The row of the
    table describes $\mathcal{L}((p, n),
    (\texttt{off\_node}, \texttt{on\_node}))$.}\label{table:example_L1}
\end{table}
Furthermore, the data global vector indices that must be sent from process $(q,
m)$ to each $(s, m) \in \mathcal{L}((q, m),
(\texttt{off\_node},\texttt{on\_node}))$ are defined as
\begin{multline}
    \mathcal{J}((q, m), (s, m), (\texttt{off\_node},
    \texttt{on\_node})) =\\
    \left\{j \in \mathcal{I}((p, n), (q, m)) \,\middle|\, A_{ij} \neq 0
    \;\textnormal{with}\; i \in \mathcal{R}((s, m))\right\}. \label{eq:IL1}
\end{multline}
The global vector indices, received from the inter-node communication step,
which $(p, n)$ must send to each local process $(q, n)$ in
Example~\ref{example:simple_matrix} are displayed in
Table~\ref{table:example_J1}.
\begin{table}[!ht]
    \centering
    \begin{tabular}{ c c c c c c c c }
       & \multicolumn{6}{ c }{$(p, n)$}\\
       & & (0, 0) & (1, 0) & (0, 1) & (1, 1) & (0, 2) & (1, 2)\\\cline{3-8}
        \multirow{6}{*}{\begin{sideways}$(q, n)$\end{sideways}}
        & \multicolumn{1}{c|}{(0, 0)}
                       & $\left\{  \right\}$
                       & $\left\{ 3 \right\}$
                       & ---
                       & ---
                       & ---
                       & --- \\
        & \multicolumn{1}{c|}{(1, 0)}
                       & $\left\{  \right\}$
                       & $\left\{  \right\}$
                       & ---
                       & ---
                       & ---
                       & --- \\
        & \multicolumn{1}{c|}{(0, 1)}
                       & ---
                       & ---
                       & $\left\{  \right\}$
                       & $\left\{  \right\}$
                       & ---
                       & --- \\
        & \multicolumn{1}{c|}{(1, 1)}
                       & ---
                       & ---
                       & $\left\{  \right\}$
                       & $\left\{  \right\}$
                       & ---
                       & --- \\
        & \multicolumn{1}{c|}{(0, 2)}
                       & ---
                       & ---
                       & ---
                       & ---
                       & $\left\{  \right\}$
                       & $\left\{ 1 \right\}$ \\
        & \multicolumn{1}{c|}{(1, 2)}
                       & ---
                       & ---
                       & ---
                       & ---
                       & $\left\{  \right\}$
                       & $\left\{  \right\}$ \\
\end{tabular}

    \caption{Global vector indices of received inter-node data that must be
    communicated between processes local to each node $n$ in
    Example~\ref{example:simple_matrix}.  Each column contains the indices of
    values sent from $(p, n)$ to $(q, n)$.  Note: dashes (---) throughout
    the table represent processes on separate nodes, which cannot communicate
    during intra-node communication.}\label{table:example_J1}
\end{table}

\subsubsection{Fully Local Communication}\label{section:L2}~\\

A subset of the values needed by a process $(p, n)$ are stored on local process
$(s, n)$.  One advantage is that these values bypass the three-step
communication, and are communicated directly.  Let $\mathcal{L}((p, n),
(\texttt{on\_node}, \texttt{on\_node}))$ define all processes local to node $n$
to which $(p, n)$ sends vector data.  This function is defined as
\begin{multline}
    \mathcal{L}((p, n), (\texttt{on\_node},
    \texttt{on\_node})) =\\
    \left\{(s, n) \,\middle|\, \exists
    \, A_{ij} \neq 0 \;\textnormal{with}\; i \in \mathcal{R}((s, n)), j \in
    \mathcal{R}((p, n)) \right\}. \label{eq:L2}
\end{multline}
The processes local to node $n$, to which $(p, n)$ must send initial vector
data in Example~\ref{example:simple_matrix} are displayed in
Table~\ref{table:example_L2}.
\begin{table}[!ht]
    \centering
    \begin{tabular}{ c c c c c c c }
                             & \multicolumn{6}{ c }{$(p, n)$}\\
                             & (0, 0) & (1, 0) & (0, 1) & (1, 1) & (0, 2) & (1,
                             2)\\\cline{2-7}
        \multicolumn{1}{c|}{$\mathcal{L}$}
        & $\left\{\right\}$
        & $\left\{(0, 0)\right\}$
        & $\left\{\right\}$
        & $\left\{(0, 1)\right\}$
        & $\left\{\right\}$
        & $\left\{\right\}$\\
\end{tabular}

    \caption{Intra-node communication requirements containing processes to which
    each process $(p, n)$ must send vector data, according to
    Example~\ref{example:simple_matrix}.  The row of the table describes
    $\mathcal{L}((p, n), (\texttt{on\_node},
    \texttt{on\_node}))$.}\label{table:example_L2}
\end{table}
Furthermore, the global vector indices that must be sent from process $(p,
n)$ to each $(s, n) \in \mathcal{L}((p, n),
(\texttt{on\_node},\texttt{on\_node}))$ is defined as follows.
\begin{multline}
    \mathcal{J}((p, n), (s, n), (\texttt{on\_node},
    \texttt{on\_node})) =\\
    \left\{j \,\middle|\, \exists
    \, A_{ij} \neq 0 \;\textnormal{with}\; i \in \mathcal{R}((s, n)), j \in
    \mathcal{R}((p, n)) \right\}. \label{IL2}
\end{multline}
The global vector indices
which $(p, n)$ must send to each local process $(s, n)$ in
Example~\ref{example:simple_matrix} are displayed in
Table~\ref{table:example_J2}.
\begin{table}[!ht]
    \centering
    \begin{tabular}{ c c c c c c c c }
       & \multicolumn{6}{ c }{$(p, n)$}\\
       & & (0, 0) & (1, 0) & (0, 1) & (1, 1) & (0, 2) & (1, 2)\\\cline{3-8}
        \multirow{6}{*}{\begin{sideways}$(s, n)$\end{sideways}}
        & \multicolumn{1}{c|}{(0, 0)}
                       & $\left\{  \right\}$
                       & $\left\{ 1 \right\}$
                       & ---
                       & ---
                       & ---
                       & --- \\
        & \multicolumn{1}{c|}{(1, 0)}
                       & $\left\{  \right\}$
                       & $\left\{  \right\}$
                       & ---
                       & ---
                       & ---
                       & --- \\
        & \multicolumn{1}{c|}{(0, 1)}
                       & ---
                       & ---
                       & $\left\{  \right\}$
                       & $\left\{ 3 \right\}$
                       & ---
                       & --- \\
        & \multicolumn{1}{c|}{(1, 1)}
                       & ---
                       & ---
                       & $\left\{  \right\}$
                       & $\left\{  \right\}$
                       & ---
                       & --- \\
        & \multicolumn{1}{c|}{(0, 2)}
                       & ---
                       & ---
                       & ---
                       & ---
                       & $\left\{  \right\}$
                       & $\left\{  \right\}$ \\
        & \multicolumn{1}{c|}{(1, 2)}
                       & ---
                       & ---
                       & ---
                       & ---
                       & $\left\{  \right\}$
                       & $\left\{  \right\}$ \\
\end{tabular}

    \caption{Global vector indices that must be communicated between
    processes local to each node $n$ in Example~\ref{example:simple_matrix}.
    Each column contains the indices of values sent
    from $(p, n)$ to $(q, n)$.  Note: dashes (---) throughout the table
    represent processes on separate nodes, which cannot communicate during
    intra-node communication.}\label{table:example_J2}
\end{table}

\subsection{Alternative SpMV Algorithm}
\begin{algorithm2e}[!ht]
  \DontPrintSemicolon	\KwIn{    \begin{tabular}[t]{l l}
        $(p, n)$ : & tuple describing local rank and\\
        & node of process\\
        $v|_{\mathcal{R}((p, n))}$: & rows of input vector $v$ local
        to\\
        & process $(p, n)$\\
        $\locality$: & locality of input and output data
    \end{tabular}
  }
  \;
	\KwOut{    \begin{tabular}[t]{l l}
        $\ell_{\text{recv}}$: & values that rank $(p, n)$ receives from\\
        & other processes\\
    \end{tabular}
  }
  \;

    \tcp{Initialize sends}
    \For{$(s, n) \in \mathcal{L}((p, n), \locality)$}{        \For{$i \in \mathcal{J}((p, n), (s, n), \locality)$} {            $\ell_{\text{send}} \leftarrow v|_{\mathcal{R}((p, n))_{i}}$\\
        }
        $\code{MPI\_Isend}(\ell_{\text{send}}, \ldots, (s, n),
            \ldots)$
    }\;
    \tcp{Initialize receives}
    $\ell_{\text{recv}} \leftarrow \emptyset$\;
    \For{$(s, n)$ s.t. $(p, n) \in \mathcal{L}((s, n), \locality)$}{        $\code{MPI\_Irecv}(\ell_{\text{recv}}, \ldots, (s, n),
            \ldots)$
    }\;
    \tcp{Complete sends and receives}
    $\code{MPI\_Waitall}$

	\caption{\code{local\_comm}}\label{alg:new_spmv_lcl}
\end{algorithm2e}

The method of communicating vector values to on-node processes is described in
Algorithm~\ref{alg:new_spmv_lcl}.
Using the definitions for the various steps of intra- and inter-node
communication, the NAPSpMV is described in Algorithm~\ref{alg:new_spmv}, where
$\code{local\_spmv}()$ refers to a row-wise, non-distributed SpMV~---~e.g.\ with
Intel's MKL library or with the Eigen Library.  It is important to note that
many slight variations to the algorithm are possible.  The fully local
communication has no dependencies, and can be performed anytime before calling
$\code{local\_spmv}(A_{\text{on\_node}}, b_{\ell \rightarrow \ell})$.
Furthermore, the function 
$\code{local\_spmv}(A_{\text{on\_process}},
v|\mathcal{R})$ has no communication requirements and, hence, can be performed
at any point in the algorithm.\\
\begin{algorithm2e}[H]
  \DontPrintSemicolon	\KwIn{    \begin{tabular}[t]{l l}
        $(p, n)$: & tuple describing local rank and node\\
        & of process\\
        $A|R$: & rows of matrix $A$ local to process (p, n)\\
        $v|R$: & rows of input vector $v$ local to process\\
        & (p, n)\\
    \end{tabular}
  }
  \;
	\KwOut{    \begin{tabular}[t]{l l}
        $w|\mathcal{R}$: & rows of output vector $w \leftarrow Av$,\\
        & local to process $(p, n)$
    \end{tabular}
  }
  \;
    $A_{\text{on\_process}} = \texttt{on\_process}(A|\mathcal{R})$\;
    $A_{\text{on\_node}} = \texttt{on\_node}(A|\mathcal{R})$\;
    $A_{\text{off\_node}} = \texttt{off\_node}(A|\mathcal{R})$\; \;

    $b_{\ell \rightarrow \ell} \leftarrow \code{local\_comm}((p, n), 
    v|\mathcal{R}, (\texttt{on\_node} \rightarrow \texttt{on\_node}))$\\

    $b_{\ell \rightarrow n\ell} \leftarrow \code{local\_comm}((p, n),
    v|\mathcal{R}, (\texttt{on\_node} \rightarrow \texttt{off\_node}))$\;
    \;

    \tcp{Initialize sends}
    \For{$(q, m) \in \mathcal{G}((p, n))$}{        \For{$i \in \mathcal{I}((p, n), (q, m))$} {            $g_{\text{send}} \leftarrow b_{\ell \rightarrow n\ell}^{i}$\;
        }
        $\code{MPI\_Isend}(g_{\text{send}}, \ldots, (q, m),
            \ldots)$
    }\;
    \tcp{Initialize receives}
    $g_{\text{recv}} \leftarrow \emptyset$\;
    \For{$(q, m)$ s.t. $(p, n) \in \mathcal{G}((q, m))$}{        $\code{MPI\_Irecv}(g_{\text{recv}}, \ldots, (q, m),
            \ldots)$
    }\;

    \tcp{Serial SpMV for local values}
    $\code{local\_spmv}(A_{\text{on\_process}}, v|\mathcal{R})$\;\;

    \tcp{Serial SpMv for on-node values}
    $\code{local\_spmv}(A_{\text{on\_node}}, b_{\ell \rightarrow \ell})$\;\;

    \tcp{Complete sends and receives}
    $\code{MPI\_Waitall}$\;\;

    $b_{n\ell \rightarrow \ell} \leftarrow \code{local\_comm}((p, n), 
    v|\mathcal{R}, (\texttt{off\_node} \rightarrow \texttt{on\_node}))$\;\;

    \tcp{Serial SpMV for off-node values}
    $\code{local\_spmv}(A_{\text{off\_node}},
        b_{n\ell \rightarrow \ell})$
	\caption{\code{NAPSpMV}}\label{alg:new_spmv}
\end{algorithm2e}

\section{Results}\label{section:results}
In this section, the parallel performance and scalability of the NAPSpMV
in comparison to the standard SpMV\@ is presented. The matrix-vector
multiplication in an algebraic multigrid (AMG) hierarchy is tested for both a
structured 2D rotated anisotropic and for unstructured linear elasticity on
$32\,768$ processes in order to expose a variety of communication patterns.  In
addition, scaling tests are considered for random matrices with a constant
number of non-zeros per row to investigate problems with no structure.  Lastly,
scaling tests on the largest 15 matrices from the SuiteSparse matrix collection
are presented.  All tests are performed on the Blue Waters parallel computer at
University of Illinois at Urbana-Champaign.

\begin{figure}[!ht]
    \centering
    \includegraphics[width=0.45\textwidth]{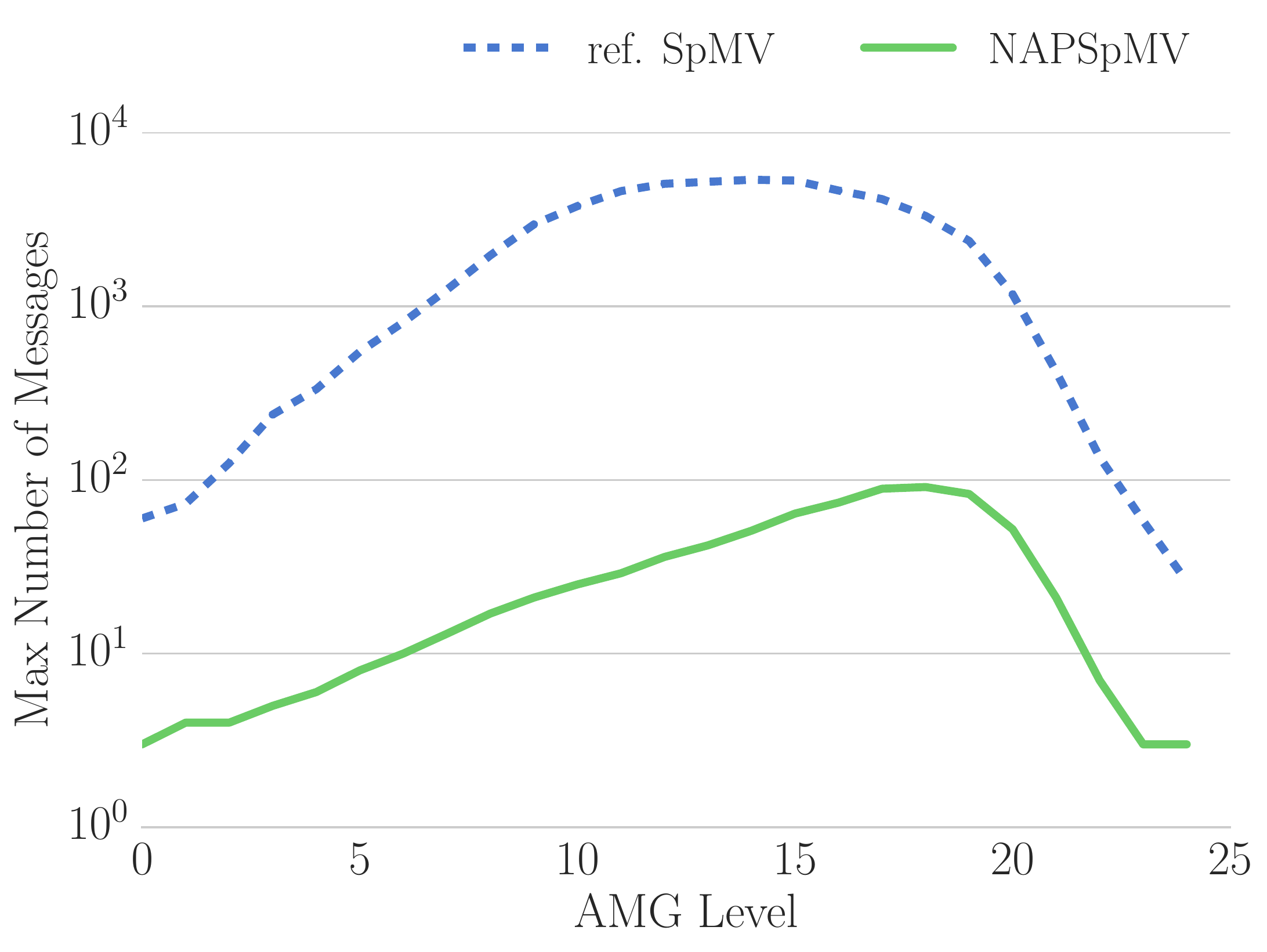}
    \includegraphics[width=0.45\textwidth]{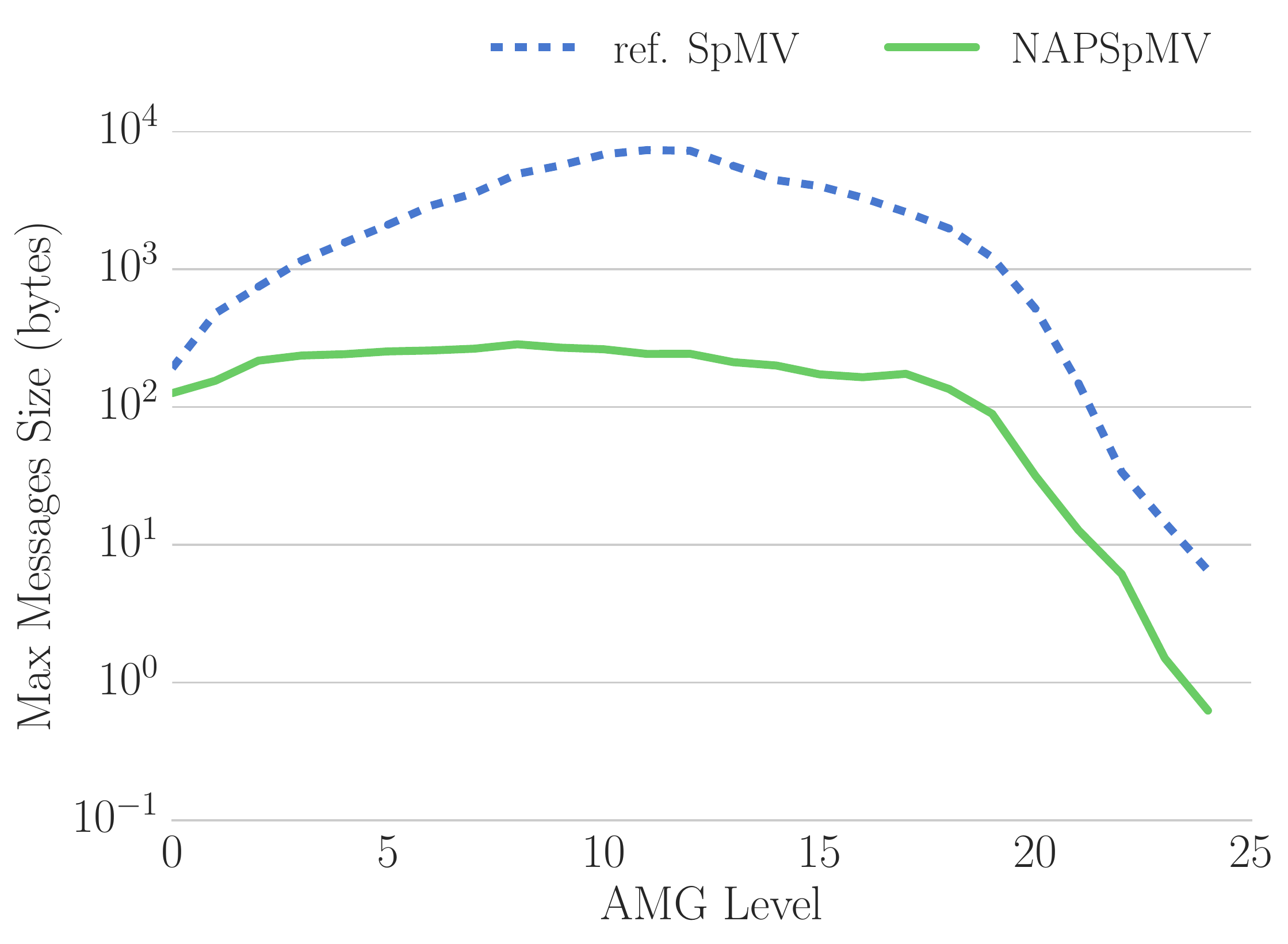}
    \caption{The maximum number (top) and size (bottom) of \textbf{inter-node}
    messages communicated by a single process during a standard SpMV and NAPSpMV
    on each level of the \textbf{linear elasticity AMG
    hierarchy}.}\label{figure:inter_comm}
\end{figure}

\begin{figure}[!ht]
    \centering
    \includegraphics[width=0.45\textwidth]{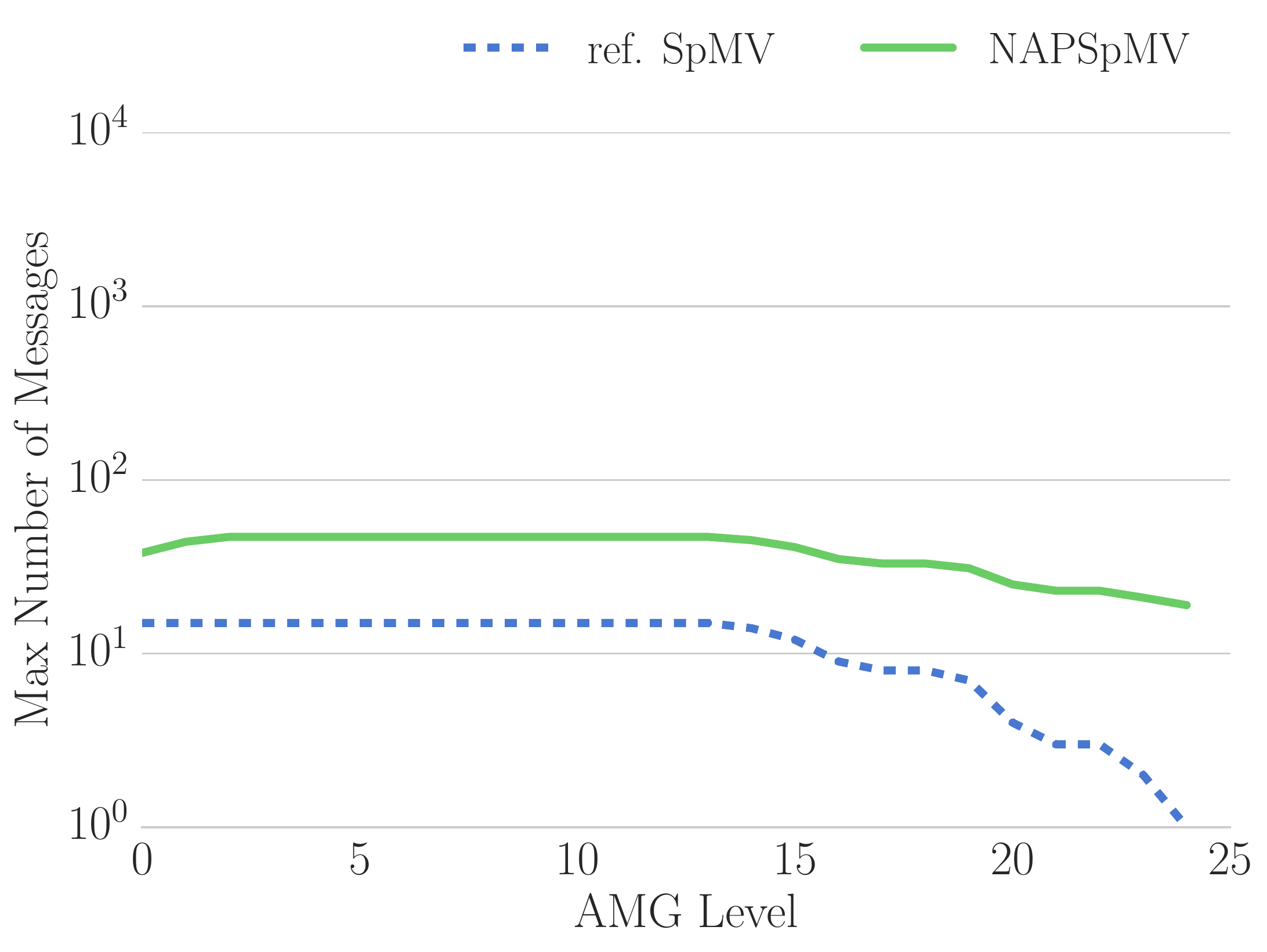}
    \includegraphics[width=0.45\textwidth]{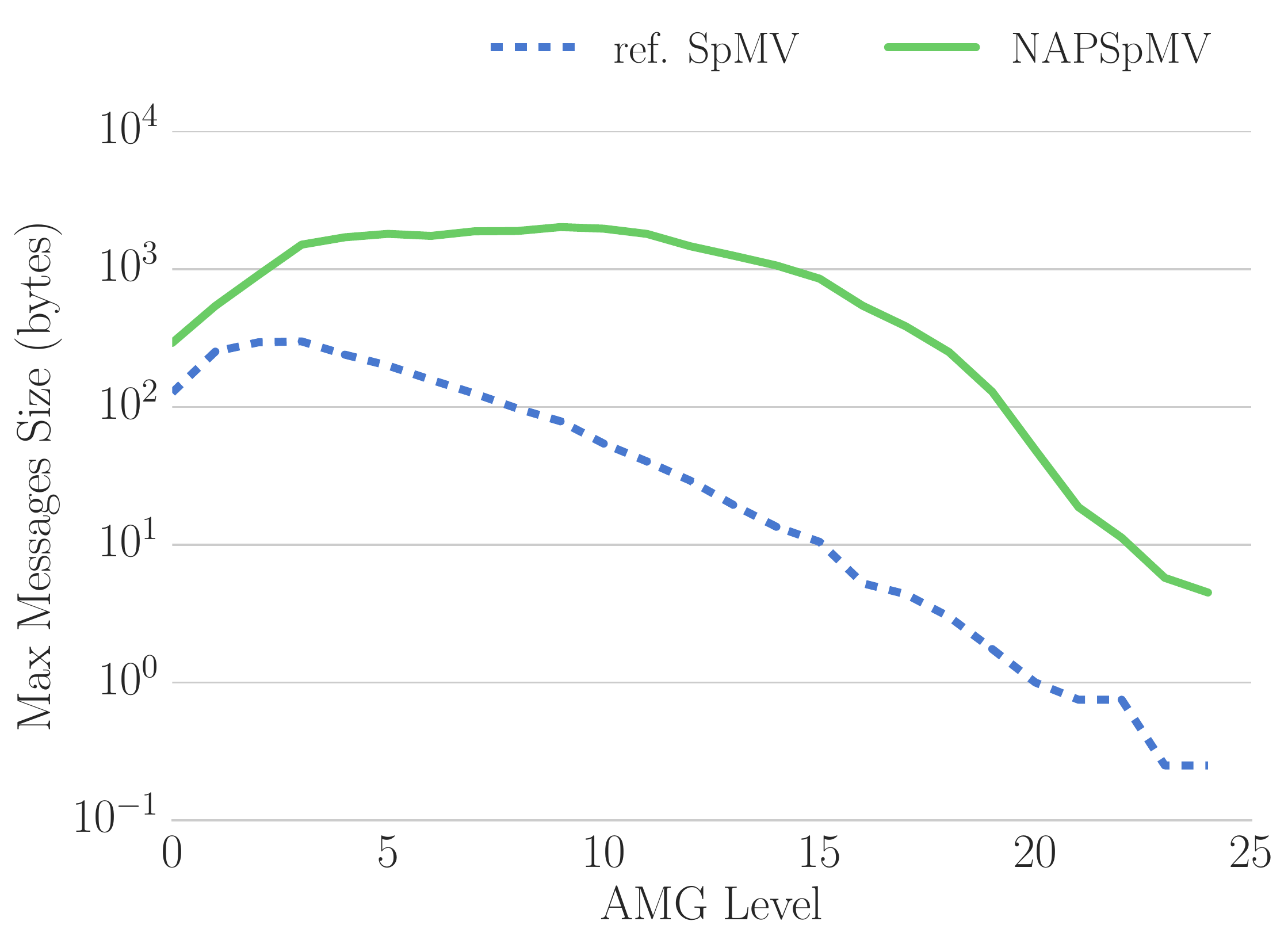}
    \caption{The maximum number (top) and size (bottom) of \textbf{intra-node}
    messages communicated by a single process during a standard SpMV and NAPSpMV
    on each level of the \textbf{linear elasticity AMG
    hierarchy}.}\label{figure:intra_comm}
\end{figure}

AMG hierarchies consist of successively coarser, but denser levels.  Therefore,
while a standard SpMV performed on the original matrix often requires
communication of a small number of large messages, coarse levels require a large
number of small messages to be injected into the network.
Figure~\ref{figure:inter_comm} shows that both the number and size of inter-node
messages required on each level of the linear elasticity hierarchy are reduced
through use of the NAPSpMV\@.  There is a large reduction in communication
requirements for coarse levels of the hierarchy, which includes a high number of
small messages.  However, as the NAPSpMV requires redistribution of data among
processes local to each node, the intra-node communication requirements increase
greatly for the NAPSpMV, as shown in Figure~\ref{figure:intra_comm}.

\begin{figure}[!ht]
    \centering
    \includegraphics[width=0.45\textwidth]{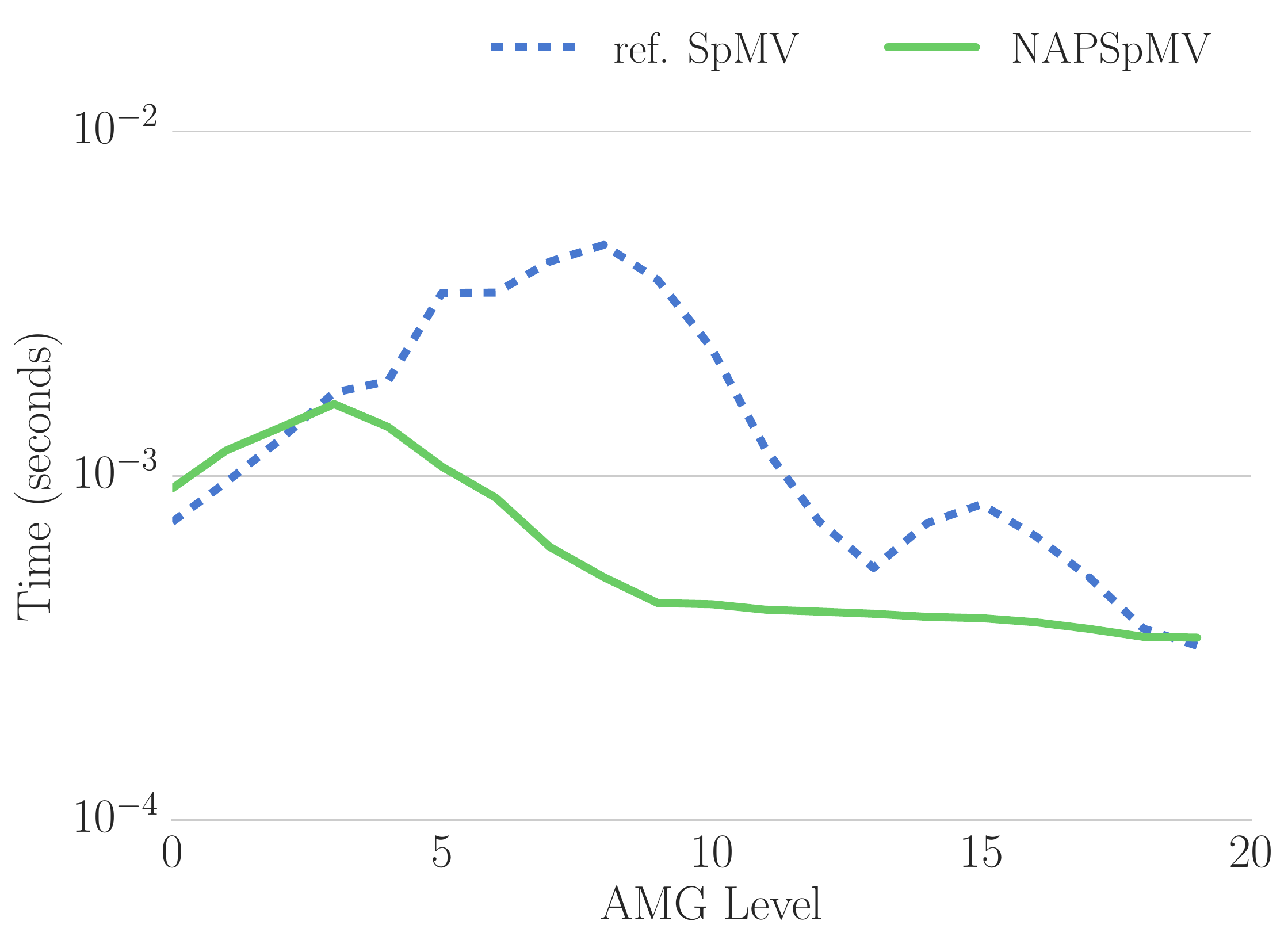}
    \includegraphics[width=0.45\textwidth]{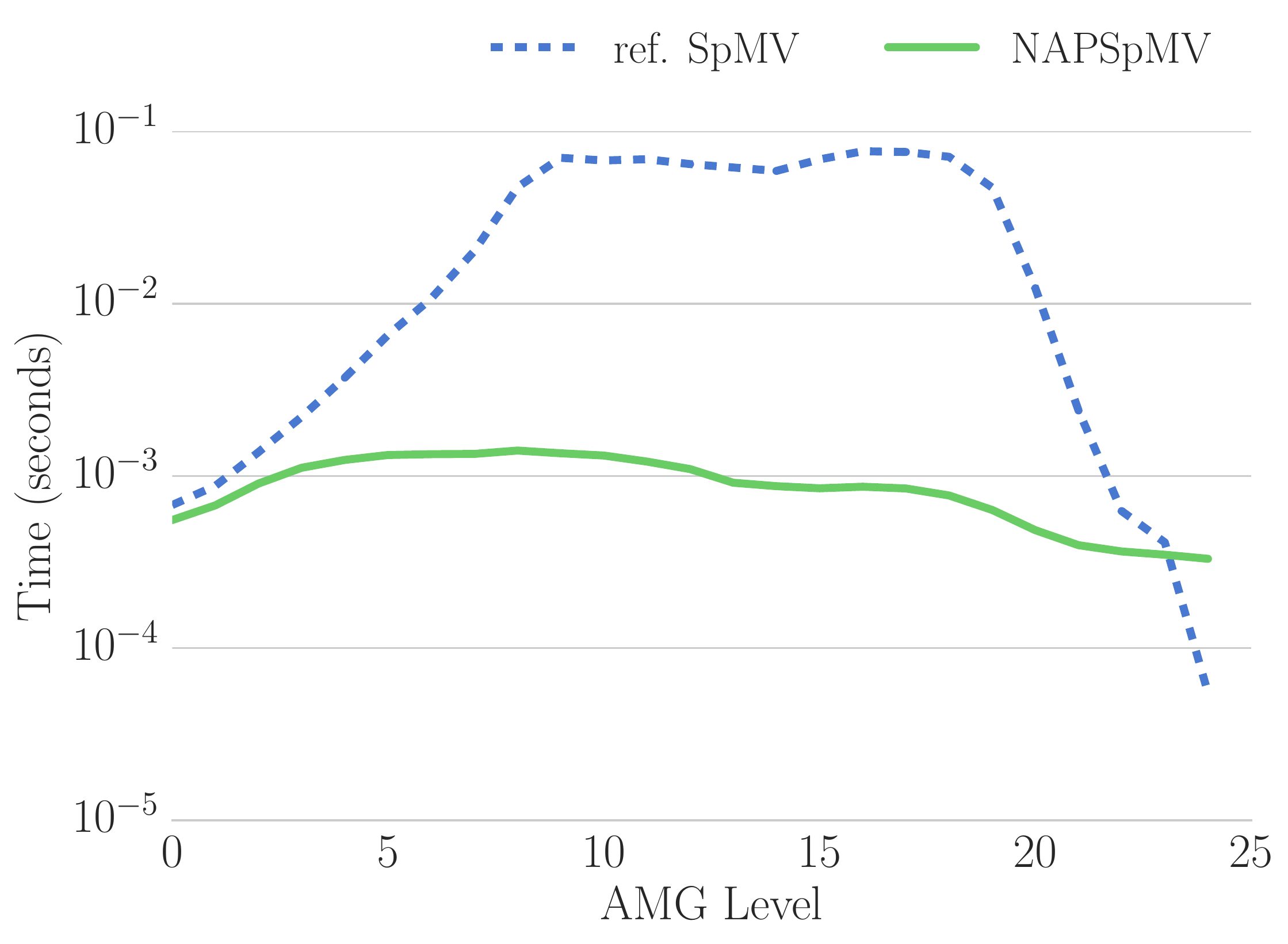}
    \caption{The time required to perform the various SpMVs on each level of the
    rotated anisotropic (left) and linear elasticity (right) AMG
    hierarchies.}\label{figure:amg_times}
\end{figure}

While there is an increase in intra-node communication requirements, the
reduction in more expensive inter-node messages results in a significant
reduction in total time for the NAPSpMV algorithm, particularly on coarser
levels near the middle of each AMG hierarchy, as shown in
Figure~\ref{figure:amg_times}.

\begin{figure}[!ht]
    \centering
    \includegraphics[width=0.45\textwidth]{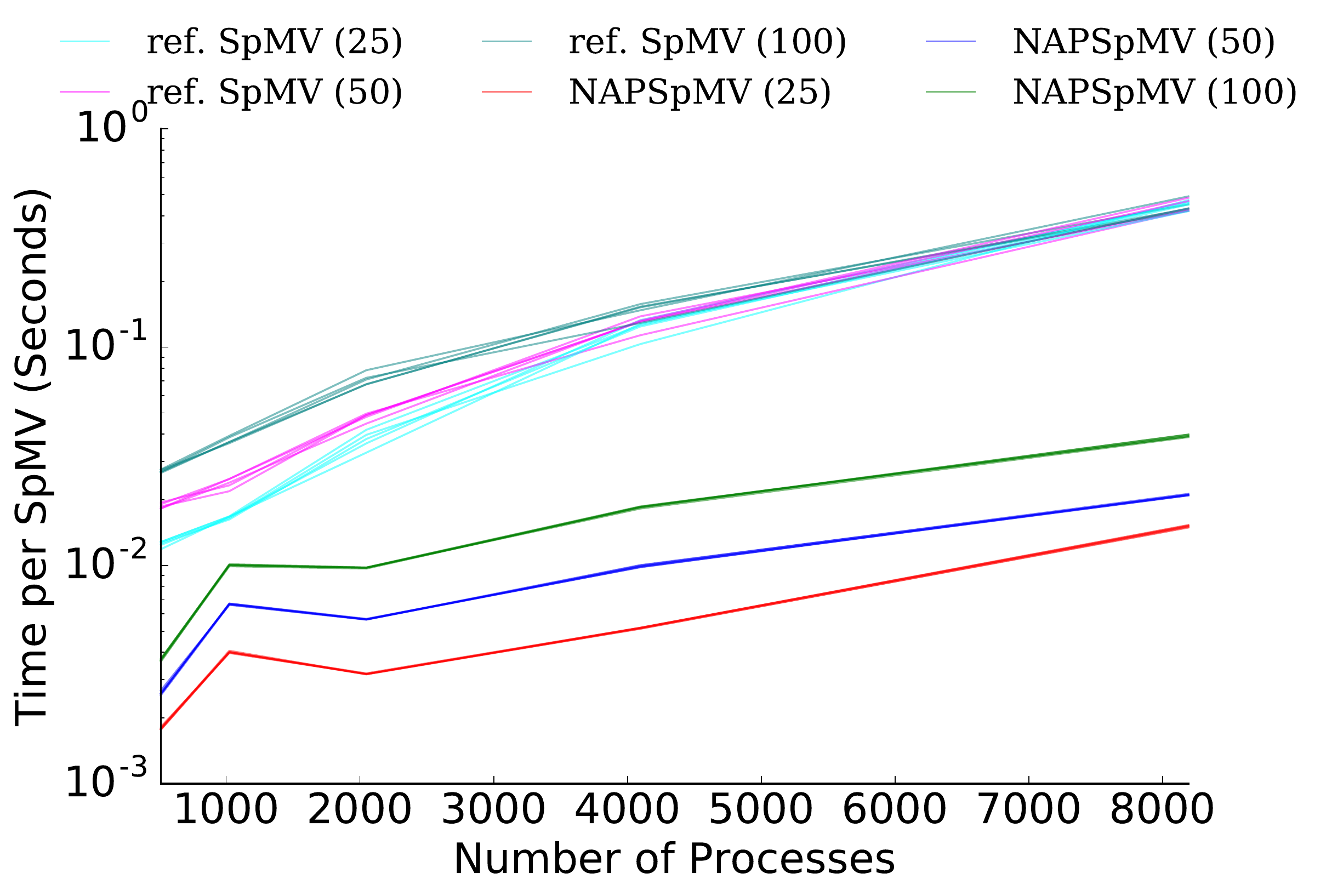}
    \includegraphics[width=0.45\textwidth]{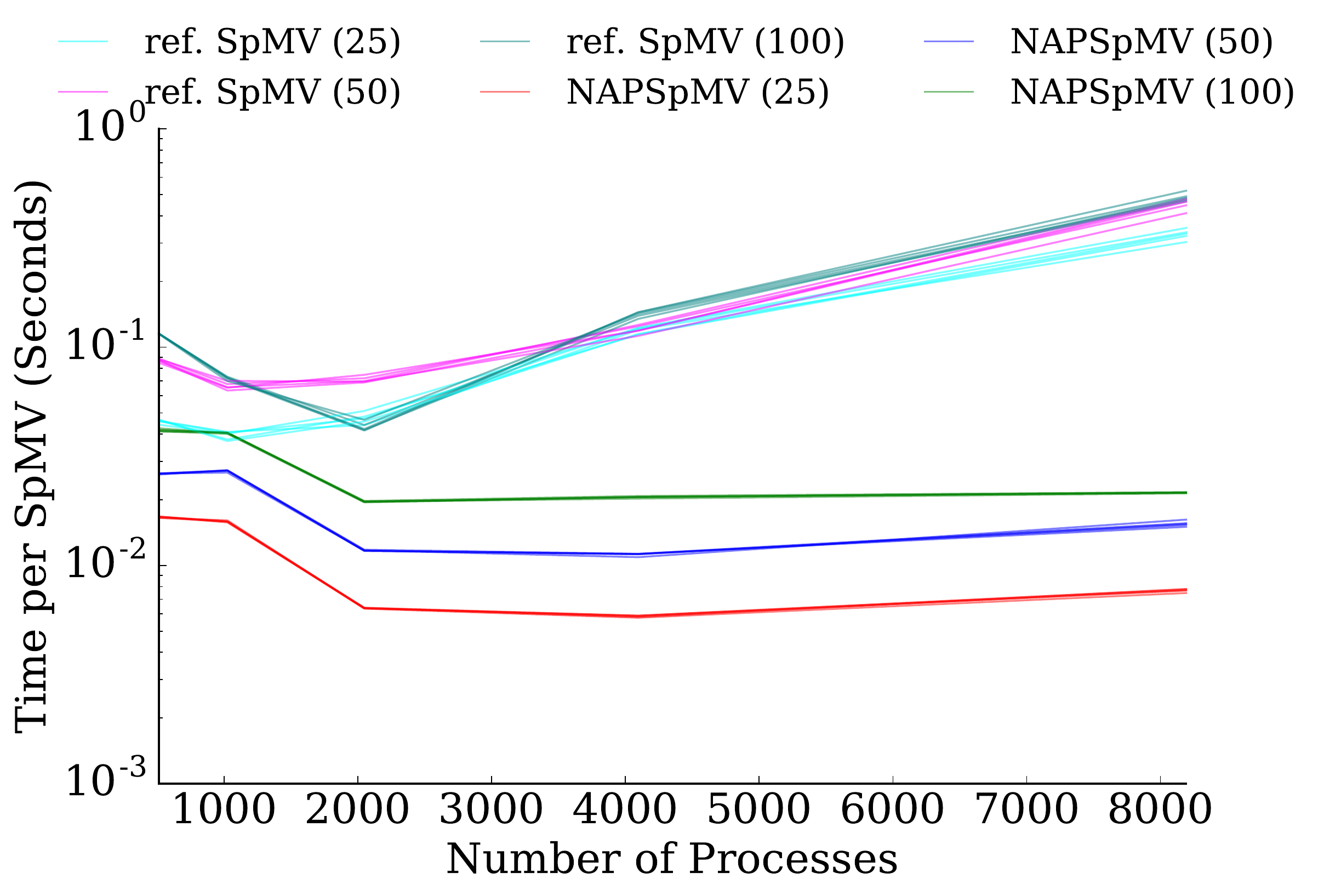}
    \caption{The time required to perform the various SpMVs on weakly (left) and
    strongly (right) scaled random matrices.  Five different random matrices are
    tested for each density of $25$, $50$, and $100$ non-zeros per row.
    The weak-scaling study tests matrices with $1\,000$ rows per process, while
    the strongly-scaled matrix contains $4\,096\,000$ rows.}\label{figure:random_compare}
\end{figure}

Random matrices, formed with a constant number of non-zeros per row, lack
structure that is found in many finite element discretizations.  As these
matrices are distributed across an increasingly large number of processes,
non-zeros are more likely to be located in off-process blocks of the matrix.
Therefore, both weak and strong scaling studies of random matrices yield
increases in communication requirements with scale.The sparsity pattern of
random matrices varies with random number generator seeds and are dependent on
the number of non-zeros per row.  Therefore, the standard SpMV and NAPSpMV were
performed on five different random matrices for each tested density of $25$,
$50$, and $100$ non-zeros per row, as shown in
Figure~\ref{figure:random_compare}.
\begin{figure}[!ht]
    \centering
    \includegraphics[width=0.45\textwidth]{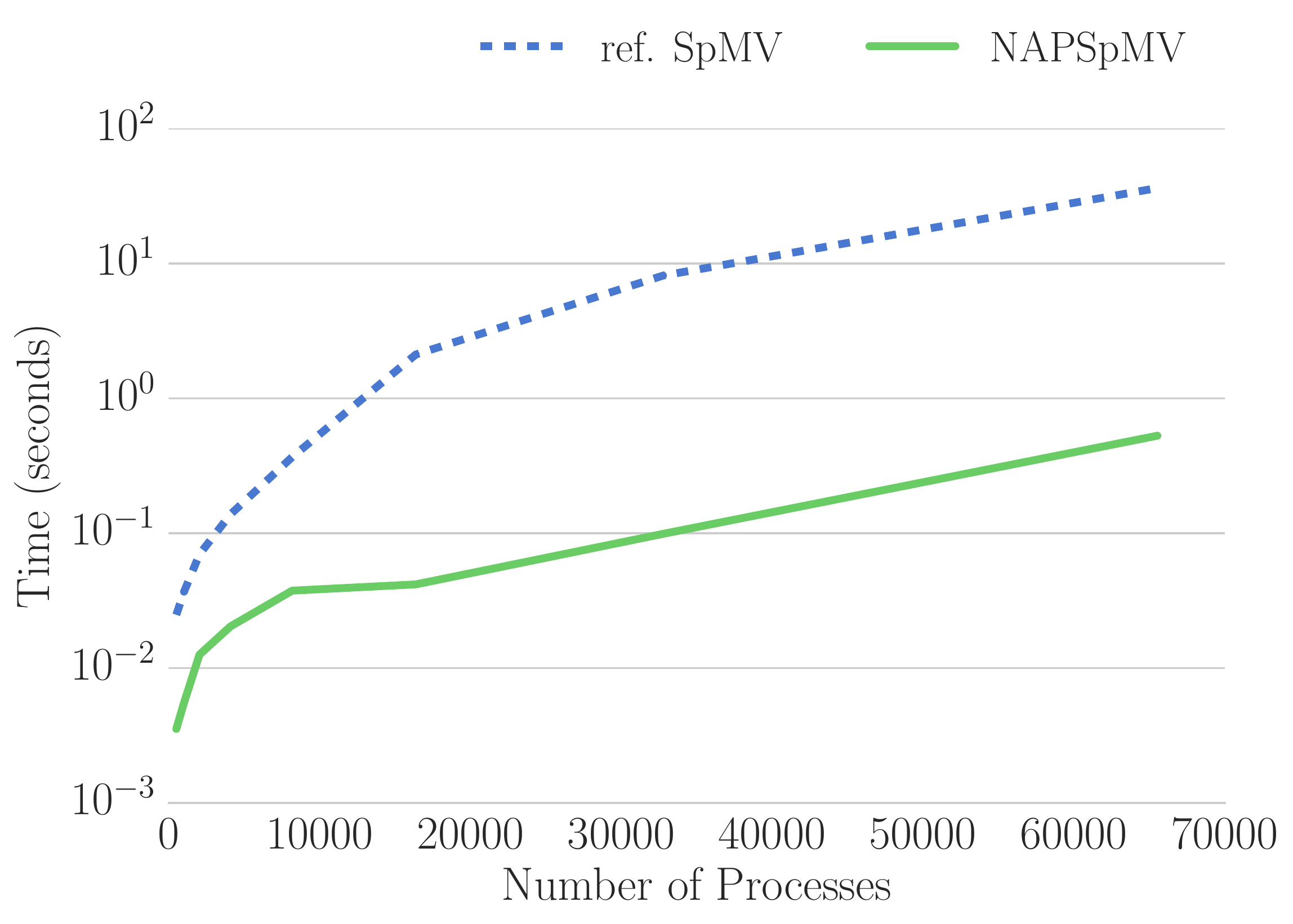}
    \includegraphics[width=0.45\textwidth]{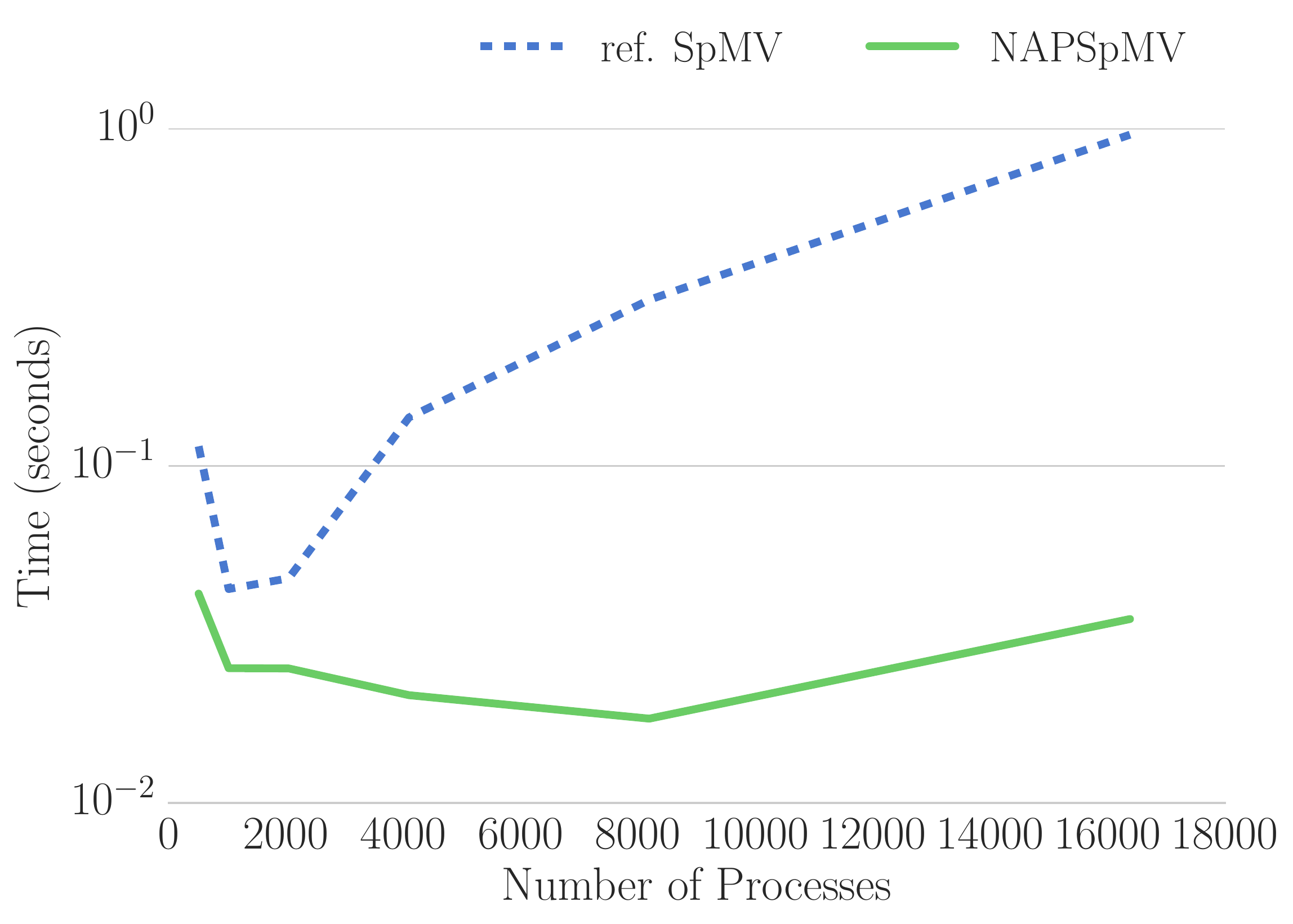}
    \caption{The time required to perform the various SpMVs on weakly (left) and
    strongly (right) scaled random matrices, each with $100$ non-zeros per row.
    The weak-scaling study tests matrices with $1\,000$ rows per process, while
    the strongly-scaled matrix contains $4\,096\,000$ rows.}\label{figure:random_times}
\end{figure}
The standard and NAPSpMV costs for all random matrices of equivalent density are
comparable.  Furthermore, there is little difference in costs between each
density.  Therefore, extended tests are performed on only a single random matrix
with $100$ non-zeros per row.  Figure~\ref{figure:random_times} displays the
time required for a NAPSpMV in comparison to the standard SpMV in both weak and
strong scaling studies.  For these random matrices, the NAPSpMV exhibits improved
performance over the reference implementation by up to two orders of magnitude and also improves
scalability.

\begin{figure}[!ht]
    \centering
    \includegraphics[width=0.9\textwidth]{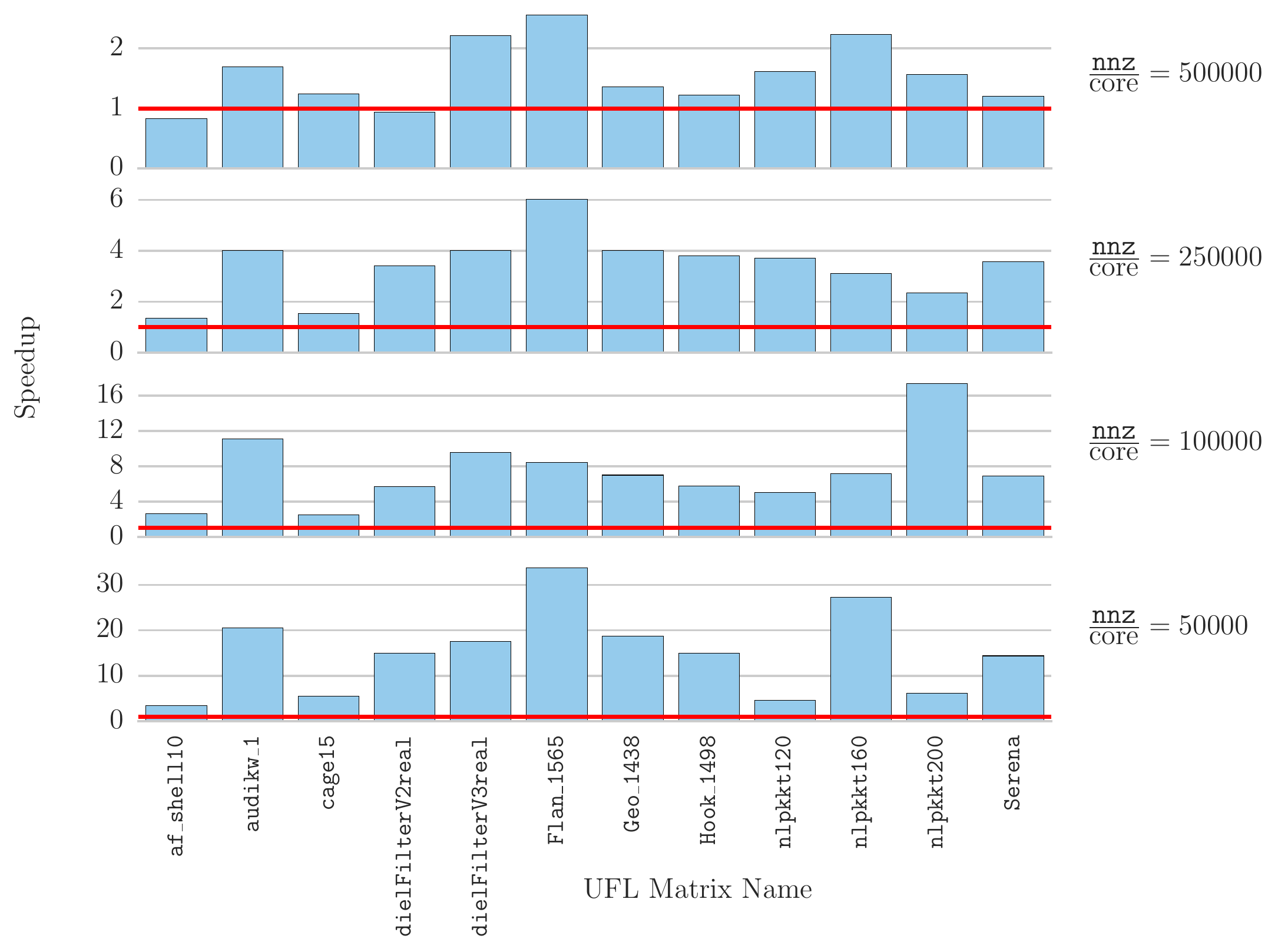}
    \caption{The speedup of NAPSpMVs over reference SpMVs on a subset of the
    largest real matrices from the SuiteSparse matrix collection at various
    scales, where $\frac{nnz}{\textnormal{core}}$ is the average number of non-zeros per core, partitioned
    so that each row $r$ is stored on process $p = r \mod n_{p}$, where $n_{p}$
    is the number of processes.}\label{figure:ufl_times}
\end{figure}

\begin{figure}[!ht]
    \centering
    \includegraphics[width=0.9\textwidth]{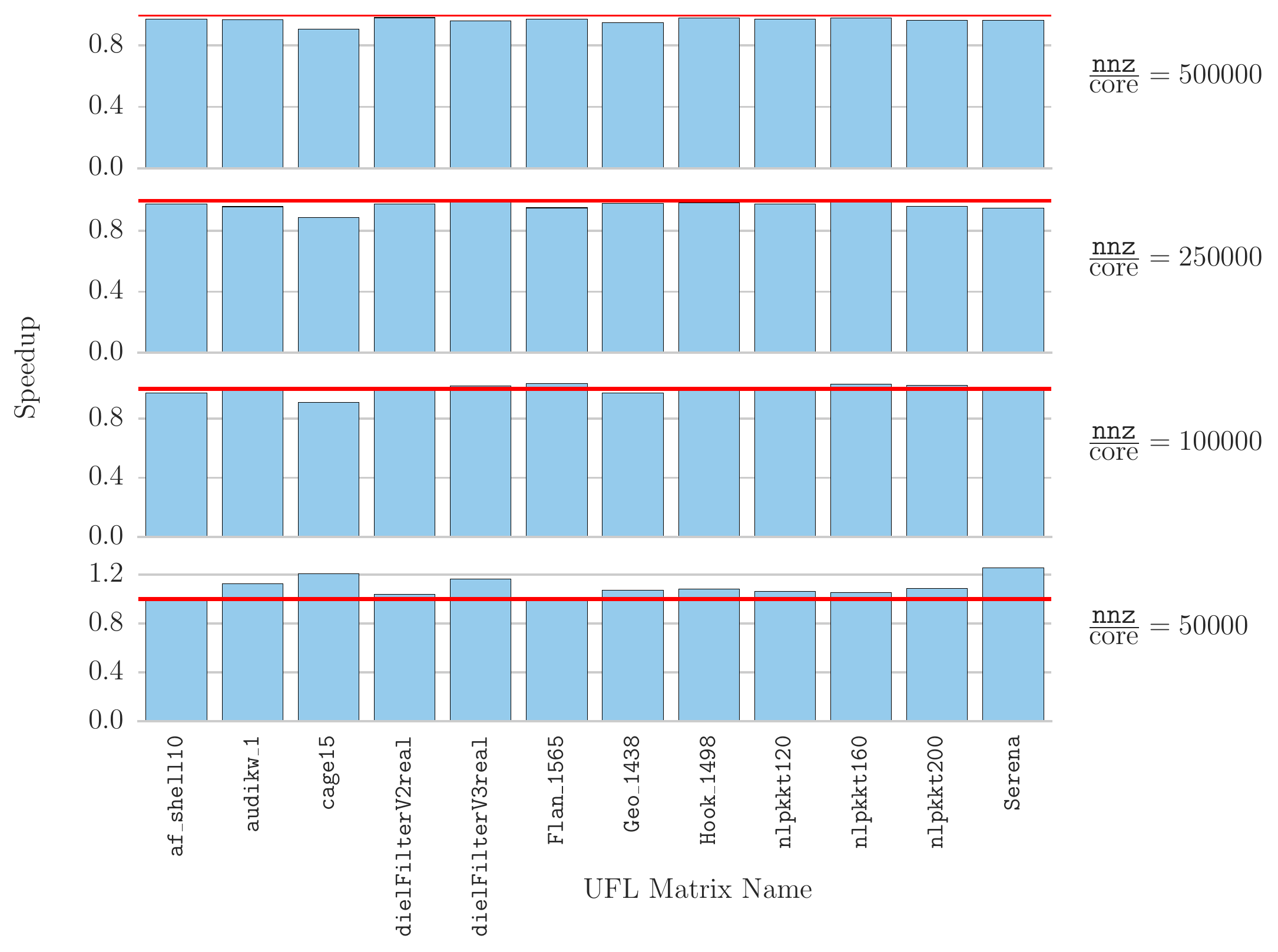}
    \caption{The speedup of NAPSpMVs over reference SpMVs on a subset of the
    largest real matrices from the SuiteSparse matrix collection at various
    scales, where $\frac{nnz}{\textnormal{core}}$ is the average number of non-zeros per core,
    partitioned with PT Scotch.}\label{figure:ufl_partitioned_times}
\end{figure}

The time required to perform the various SpMVs on $13$ of the $15$ largest
matrices from the SuiteSparse matrix collection are shown with strided and
balanced partitions, in
Figures~\ref{figure:ufl_times}~and~\ref{figure:ufl_partitioned_times}
respectively.  The remaining $2$ large matrices were not included due to
partitioning constraints.  
For the strided partitions with $n_{p}$ processes,
each row $r$ is local to process $p = r \mod n_{p}$.  As some matrices in this
subset have nearly dense blocks of rows, this allows for improved load balancing
over each process holding a contiguous block of rows.  The balanced partitions
were formed with PT Scotch graph partitioning, using the strategy
\code{SCOTCH\_STRATBALANCE}.

\begin{figure}[!ht]
    \centering
    \includegraphics[width=0.9\textwidth]{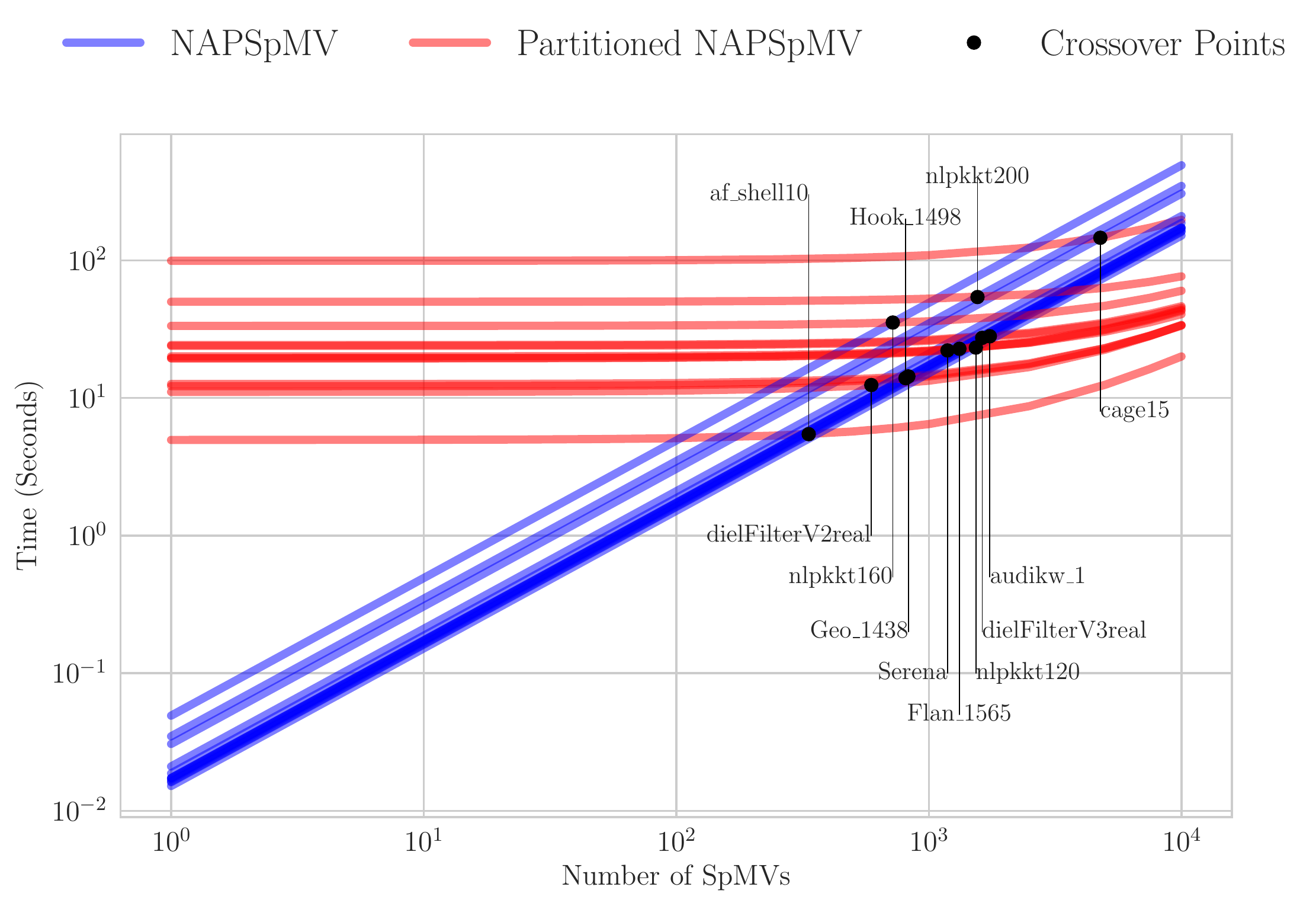}
    \caption{The time required to perform various numbers of NAPSpMVs on
    strided and balanced partitions of the largest real SuiteSparse matrices with
    $50\,000$ non-zeros per process.  The time to perform a NAPSpMV on a
    balanced partition includes the setup cost of partitioning and
    redistributing the matrix.  The crossover points represent the number of
    NAPSpMVs required before graph partitioning becomes less costly than
    performing NAPSpMVs on the strided partition.}\label{figure:partition_times}
\end{figure}

The NAPSpMV improves upon many of the matrices with strided partitions, as
communication patterns are far from optimal, while only minimally improving upon
the graph partitioned matrices.  However, the cost of partitioning motivates the
use of less optimal partitions when a smaller number of SpMVs are to be
performed.  Figure \ref{figure:partition_times} shows the time required to
perform various numbers of NAPSpMVs on both the strided and balanced partitions
at the strongest scale tested, with $50\,000$ non-zeros per core.  
In these tests, the balanced partitioned timings include the time required to
graph partition and redistribute the matrix. The crossover point for the various
SuiteSparse matrices, at which the graph partitioning becomes less costly than
performing NAPSpMVs on strided partitions, occurs only after hundreds, or
often thousands, of SpMVs have been performed.

\section{Conclusion and Future Work}\label{section:conclusion}

\begin{figure}[!ht]
    \centering
    \includegraphics[width=0.6\textwidth]{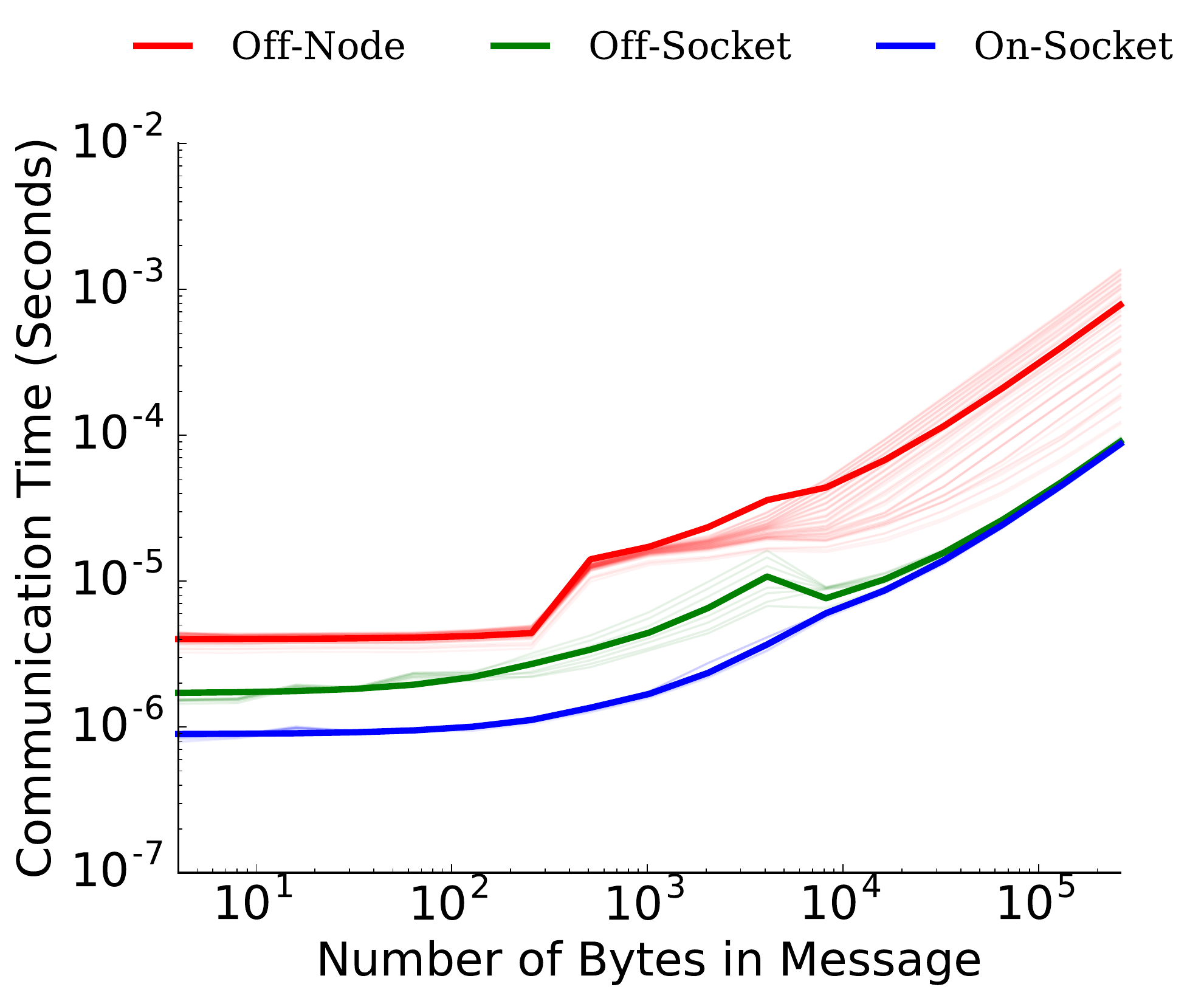}
    \caption{The time required to send a single message of various sizes,
    with the thin lines representing timings measured by Nodecomm and the thick
    lines displaying the \textit{max-rate} and intra-node models
    in~\eqref{eq:max_rate}~and~\eqref{eq:intra_comm},
    respectively.  The intra-node models are split into two categories,
    on-socket and off-socket.}\label{figure:model_msg_times_separated}
\end{figure}

This paper introduces a method to reduce communication that is injected into the
network during a sparse matrix-vector multiply by reorganizing messages on each
node.  This results in a reduction of the inter-node communication, replaced by
less-costly intra-node communication, which reduces both the number and size of
messages that are injected into the network.  The current implementation could
be extended to take various levels of the hierarchy into account, such as
splitting intra-node messages into on-socket and off-socket.
Figure~\ref{figure:model_msg_times_separated} shows that on-socket messages are
significantly cheaper and could be targeted to further reduce communication
costs.

\bibliographystyle{elsarticle-num}
\bibliography{paper}

\end{document}